\documentclass[%
 reprint,
 amsmath,amssymb,
 aps,
]{revtex4-2}

\usepackage{graphicx}
\usepackage{dcolumn}
\usepackage{bm}

\usepackage{gensymb}
\usepackage{xcolor}
\usepackage{soul}

\usepackage{mathtools}
\DeclarePairedDelimiter{\abs}{\lvert}{\rvert}
\DeclarePairedDelimiter{\aver}{\langle}{\rangle}

\usepackage{bm}
\renewcommand{\vec}[1]{\bm{#1}}
\renewcommand{\vec}[1]{\bm{#1}}

\usepackage{upgreek}
\renewcommand{\mu}{\upmu}
\renewcommand{\textcolor}[2]{#2}

\begin{document}

\preprint{APS/123-QED}

\title{Silicon Carbide photonic platform based on suspended subwavelength waveguides}

\author{Francesco Garrisi}
\email{francesco.garrisi01@universitadipavia.it}
\affiliation{%
 Dipartimento di Fisica,
 Università di Pavia,
 via A. Bassi 6, 27100 Pavia, Italy
}%
 
\author{Ioannis Chatzopoulos}%
 \altaffiliation[Currently at ]{National Physical Laboratory, Hampton Road, Teddington TW11 0LW, United Kingdom}

\author{Robert Cernansky}
 \altaffiliation[Currently at ]{Centre for Quantum Dynamics, Grifﬁth University, Brisbane QLD 4111, Australia}
 
\author{Alberto Politi}
\affiliation{%
 School of Physics and Astronomy, University of Southampton, Southampton, SO17~1BJ, United Kingdom
}%


\begin{abstract}
Silicon carbide (SiC) displays a unique combination of optical and spin-related properties that make it interesting for photonics and quantum technologies.
However, guiding light by total internal reflection can be difficult to achieve, especially when SiC is grown as thin films on higher index substrates, like silicon.
Fabricating suspended subwavelength waveguides requires a single lithography step and offers a solution to the confinement problem, while preserving the design flexibility required for a scalable and complete photonic platform.
Here we present a design for such platform, \textcolor{red}{which} can be used for both classical and quantum optics operation. We simulate \textcolor{red}{basic} optical components and analyze how to exploit the high nonlinearities of SiC and its defects.

\end{abstract}

\maketitle

\section{Introduction}
\label{sec:introduction}
Silicon carbide (SiC) is establishing itself as an important material in the field of quantum photonics. Among its many polytypes, 3C and 4H-SiC are hosts of a large variety of point defects emitting in the visible and in the near infrared (NIR) \cite{koehl2011room, falk2013polytype, castelletto2014silicon, widmann2015coherent}; these defects can be used as single photon sources and their spin state can be addressed through radio frequency and optical electromagnetic fields, while the coherence time has been shown to exceed milliseconds \cite{christle2015isolated, christle2017isolated, simin2017locking}.

Photonic structures can enhance the interaction between these colour centres and light, providing a path for the development of a scalable approach for quantum technologies.
Moreover, SiC provides interesting optical properties.
Being non-centrosymmetric crystals, both polytypes of SiC possess a strong static second-order nonlinearity ($\chi^{(2)}_\text{xyz} \simeq 60\ \text{pm/V}$ for 3C-SiC \cite{tang1991linear} and $\chi^{(2)}_\text{zzz} \simeq 32.8\ \text{pm/V}$ for 4H-SiC \cite{wu2008second}); they do not suffer from two-photon absorption at telecommunication wavelengths due to their large electronic bandgap (around 2.4~eV for 3C and 2.9~eV for 4H \cite{madelung1982physics}); as well as diamond, SiC is one of the hardest known materials \cite{jackson2005mechanical}, providing the mechanical stability required to support complex nanostructures at small scale \textcolor{red}{\cite{chatzopoulos2019high}}, along with excellent  thermal conductivity.
\textcolor{red}{SiC is also being investigated as a dielectric photonic platform for plasmonic applications \cite{caldwell2013low}.}
Finally, SiC is known to be an established platform for high power microelectronics, making promising the integration of photonic and electronic devices on the same platform.


The fabrication of SiC for photonic applications, however, can be problematic.
For example, few hundred nanometers of 3C-SiC can be grown heteroepitaxially on silicon (Si), but this poses two issues:
i) having a higher index of refraction, the substrate prevents the use of total internal reflection (TIR) to obtain light confinement in the vertical direction;
ii) due to crystalline mismatch, the interface between 3C-SiC and Si grows with very low quality, increasing losses of light travelling in such region. These two problems can be addressed at once by adopting wafer-bonding techniques \cite{fan2018high}.
On the other hand, the homoepitaxial growth of 4H-SiC provides high quality films, but obtaining thin membranes is not straightforward.
Smart-cut process \cite{di1996silicon} can be applied to obtain SiC on insulator, but the ion implantation step increases optical losses and produces lattice damages detrimental to color center properties.
Wafer bonding and thin down has demonstrated excellent material properties and low losses in photonic crystal cavities \cite{Song2019ultrahigh} and ring resonators \cite{lukin20194h}. However, the uniformity of the thickness of the SiC layer over appreciable chip sizes is a limiting factor for the scalability of SiC photonics.

An alternative approach is to suspend membranes in air, either removing part of the Si substrate, or by electrochemical etching of doped SiC.
This approach has been used to produce photonic crystal cavities \cite{bracher2015fabrication, calusine2014silicon} as well as optical waveguides using a two-step lithography technique \cite{martini2017linear}. In this case, the first \textcolor{red}{exposure} defines the lateral confinement of the waveguides, while the second one opens holes to access the substrate that has to be removed.



Here we propose subwavelength geometries that allow the use of a single \textcolor{red}{lithography} step to \textcolor{red}{define the waveguide geometry and} access the substrate, simplifying the fabrication, as it has already been demonstrated for other platforms \cite{penades2014suspended,Penades16,penades2018suspended,zhou2018fully}.
Subwavelength structures can be defined as periodic dielectric structures whose periodicity is much smaller than the wavelength of light; more rigorously, they are periodic structures where the energy of the photonic bandgap lies above the energy of the photons propagating in the medium. As such, they behave as an effective homogeneous medium (EHM) and they prevent the scattering of light \cite{cheben2018subwavelength}.
Subwavelength structures can be used easily to obtain a complete photonic platform in SiC, capable not only to guide light, but also to realize ring resonators, grating couplers and slow-light waveguides.\\

In Section \ref{sec:waveguide} the design of the most basic component of the platform, a straight subwavelength suspended waveguide, is presented, followed by a discussion on the results of the numerical simulations which led to the choice of the dimensions of the waveguide.
Then, we consider the amount of losses expected from the waveguide design, and we give an estimate of the nonlinear waveguide parameter.
In Section \ref{sec:tolerance} we assess the tolerance of the design to fabrication imperfections, in terms of the variation of the modal refractive index resulting from variation in the geometry of the waveguide.
Section \ref{sec:devices} briefly introduces the analysis of additional photonic components and presents detailed results of numerical simulations used to design a uniform grating coupler.
Section \ref{sec:slowlight} describes how the platform can be adapted easily to reach a slow-light regime by changing the periodicity of the lateral suspending structures.
Section \ref{sec:modulators} discusses the performances of a proposed design for an electro-optical modulator integrated alongside the suspended waveguides.
Finally, in Section \ref{sec:conclusions} we give the conclusions and perspectives.

\section{Subwavelength Waveguide}
\label{sec:waveguide}

\begin{figure*}
    \centering
    \includegraphics[width=0.8\textwidth]{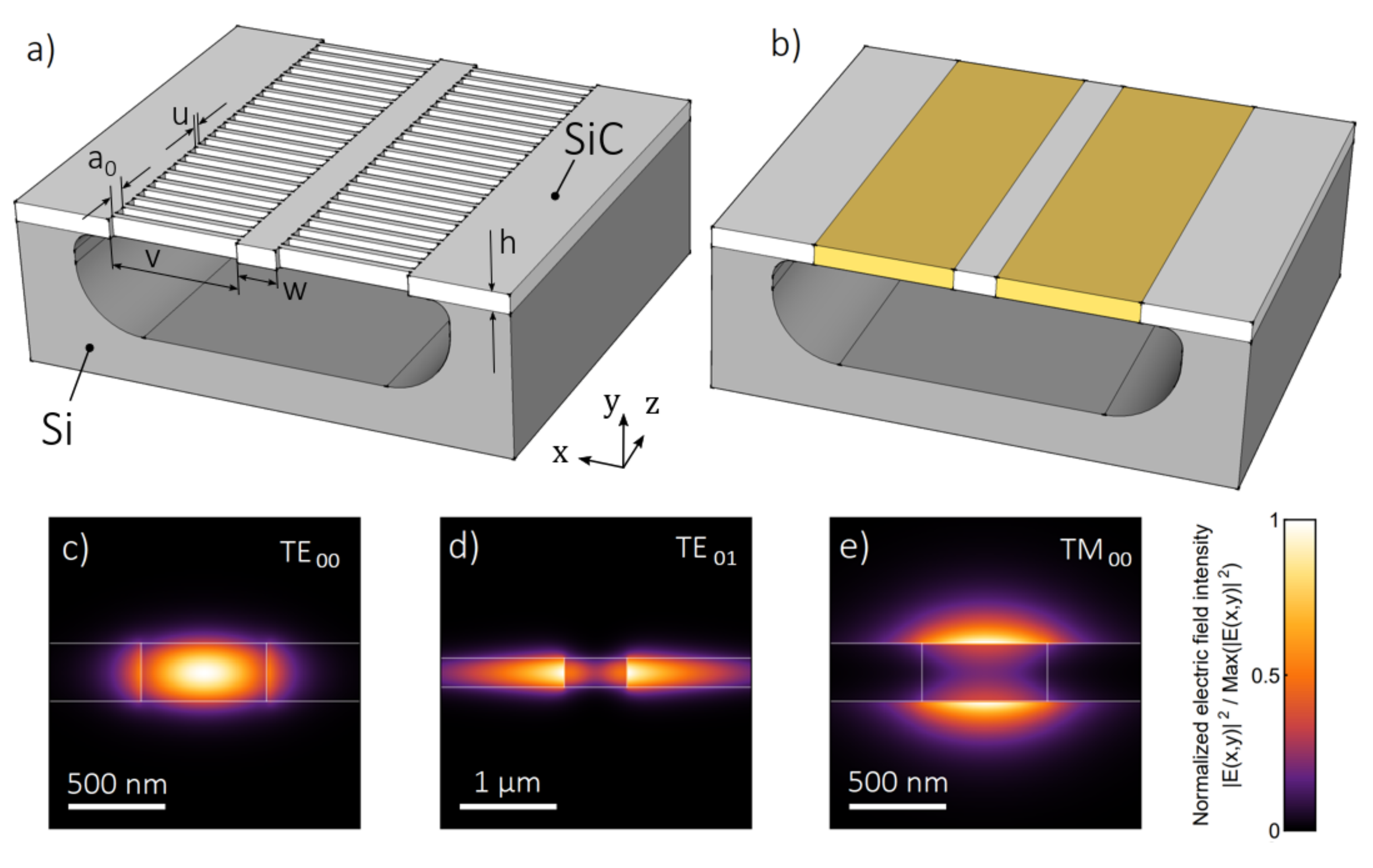}
    \caption{a) Scheme of the proposed design of a SiC suspended waveguide. The SiC film is 300~nm thick and sits on top of a Silicon substrate. The central branch (650~nm wide) is supported by the lateral arms, which have a longitudinal periodicity of 300~nm and dimensions of 2~$\mu$m $\times$ 150~nm. b) The arms act as an effective uniform medium (yellow) and contribute to guide light by total internal reflection. c-e) Electric field intensity profiles of the $\text{TE}_{00}$, $\text{TE}_{01}$ and $\text{TM}_{00}$ modes.}
    \label{fig:WGscheme}
\end{figure*}

When a dielectric medium is periodic in one direction, the light travelling inside it can be described in terms of the photonic band structure \cite{bookjoannopoulos}.
The periodicity will produce the emergence of the photonic bandgap, a range of frequencies at which light cannot propagate in the medium.
If the energy of the light is lower than the photonic bandgap, radiation can propagate, ideally without scattering, and the periodic medium acts as an EHM \cite{cheben2018subwavelength}.

In Figure \ref{fig:WGscheme}-a) we show the design of a SiC waveguide that exploits this principle to guide light at 1550 nm wavelength.
The design is based on previous works realized in silicon on insulator (SOI) \cite{Penades16,penades2018suspended}.
Light is confined in the vertical direction by TIR.
The lateral arms serve two functions: to mechanically suspend the waveguide and to introduce the periodic perturbation.
The perturbation has a periodicity that is much smaller than the wavelength of light, hence, similarly to a multilayer, the arms act as a homogeneous medium with index of refraction $n_e$ intermediate between the one of SiC and air. Thus, for the case of the straight waveguide presented here, the structure is akin to the one shown in Figure \ref{fig:WGscheme}-b), where the yellow region highlights the EHM; in practice, this confines light by TIR on the horizontal direction, as well.


The bulk effective index $n_e$ of the subwavelength region can be tuned changing the filling factor (FF) of the arms $f_\text{wg}$ in the periodic cell, and can be estimated by calculating the effective index of the light travelling normal to a SiC-air multilayer with the same periodicity and FF of the lateral arms \cite{yariv2006photonics}.
The minimum feature size given by the fabrication process sets the constraints for $f_\text{wg}$ and hence to $n_e$.
For our SiC structure we believe the higher limit on $f_\text{wg}$ will be set by the resolution of the lithographic process, while the lower limit will be determined by the mechanical strength of the material.
For instance, other structures in SOI \cite{Penades16,penades2018suspended} were fabricated with a minimum dimension of the arms equal to 100~nm.
Since SiC is a very hard material \cite{jackson2005mechanical}, it is reasonable to assume that the minimum dimension of the arms could be smaller than this value, but a more detailed analysis is required that takes into account not only the mechanical stability of the material but also the inner stress.


\begin{table}
    \centering
    \begin{tabular}{c c}
    \hline
    \text{Dimension} & \text{Length [nm]} \\
    \hline
    \text{Film thickness ($h$)} & 300 \\
    \text{Waveguide width ($w$)} & 650 \\
    \text{Periodicity} $(a_0)$ & 300 \\
    \text{Arm width ($u = f_\text{wg}\,a_0$)} & 150 \\
    \text{Arm length ($v$)} & 2000 \\
    \hline
    \end{tabular}
    \caption{Proposed dimensions for a single TE-TM subwavelength waveguide. We assume a value of 2.6 for the index of refraction of SiC, suspension in air and vertical walls.}
    \label{tab:dimensions}
\end{table}

\begin{figure}
    \centering
    \includegraphics[width=0.45\textwidth]{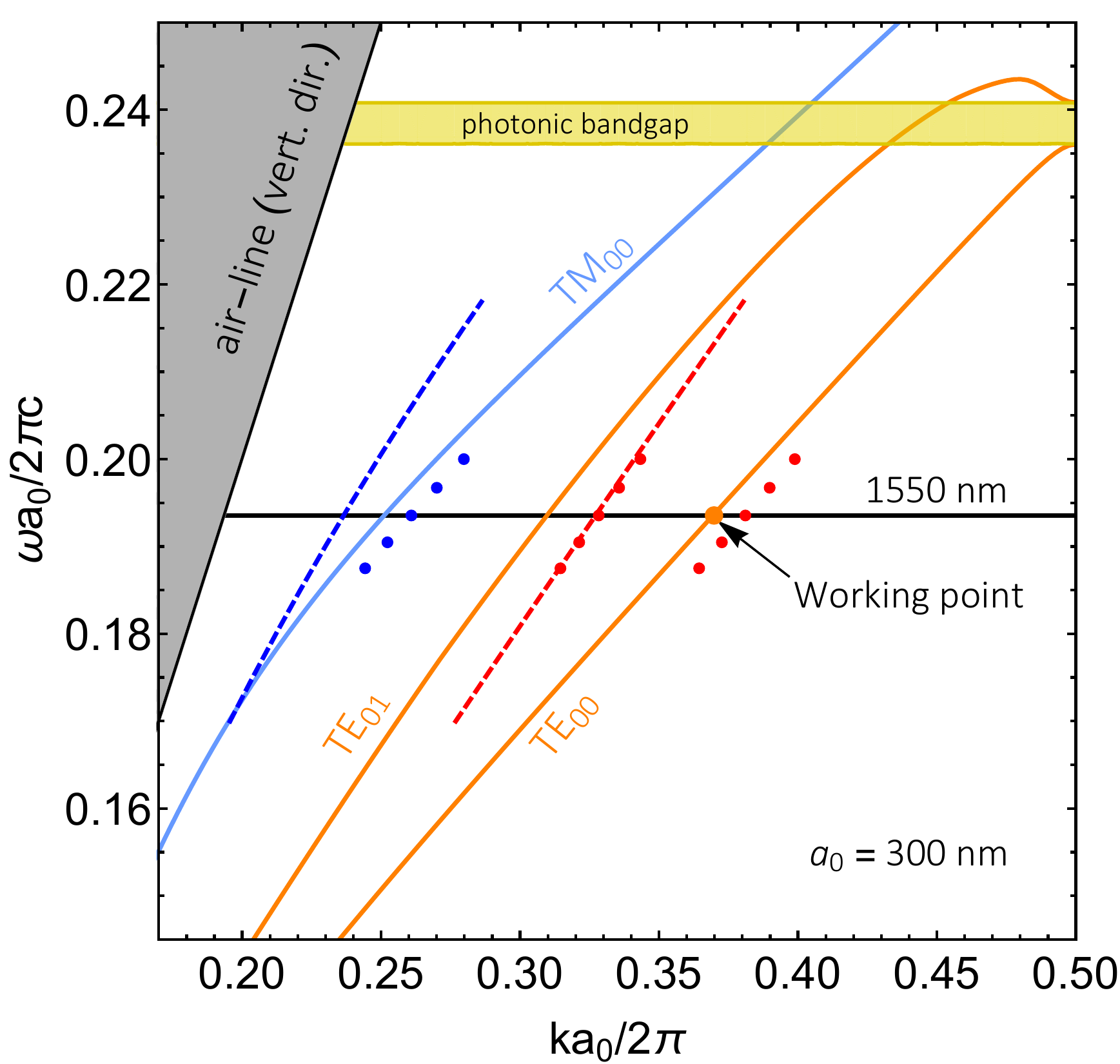}
\caption{Simulated band structure of the proposed suspended waveguide along the propagation direction. Assuming \textcolor{red}{$a_0$} = 300~nm, the horizontal black line corresponds to 1550~nm. Orange and light blue lines correspond to guided TE and TM modes respectively, calculated with the MIT MPB simulation suite \cite{johnson2001block}; red and blue dots are the same modes calculated with a numerical eigensolver; dashed red and blue lines are effective horizontal TE and TM light-lines calculated from an effective index approach.}
    \label{fig:Bands300}
\end{figure}

\textcolor{red}{The goal of the design process is to obtain a structure sustaining a single TE mode, while maximizing its confinement.}.
The dimensions of the proposed structure are listed in Table \ref{tab:dimensions}.
We assumed a value of 2.6 for the refractive index of the SiC layer since it is close to the refractive indexes of both 3C and 4H-SiC \textcolor{red}{\cite{madelung1982physics}}.
Then, the layer thickness ($h$) of 300 nm was chosen in order to have the fundamental slab mode close to the cut-off condition.
The periodicity of the structure ($a_0$) is set by the subwavelength condition:
the continuous lines of Figure \ref{fig:Bands300} are the band structure of our subwavelength waveguides calculated using MIT Photonic Bands (MPB) \cite{johnson2001block};
choosing a periodicity of 300~nm puts the TE bandgap well above the energy of 1550~nm radiation, ensuring the suppression of scattered light and the validity of the EHM approximation.

We have chosen \textcolor{red}{$f_\text{wg} = u/a_0$ equal to 0.5, so that the lateral arms are 150 nm long ($u$) in the propagation direction.
As explained previously, this corresponds to having} $n_e$ equal to 2.144, a good compromise between a higher lateral confinement and a high mechanical strength.
The proposed waveguide's width ($w$) of 650 nm is the one that maximizes the confinement of the fundamental mode while maintaining the structure single-TE-moded.
In fact, the structure sustains a single TM mode ($\text{TM}_{00}$) and two TE modes ($\text{TE}_{00}$ and $\text{TE}_{01}$);
the $\text{TE}_{01}$ mode is very loosely bound and is likely to experience very high losses compared to the $\text{TE}_{00}$ mode, since it would be easily coupled to radiative modes.
\textcolor{red}{The effective mode area of the $\text{TE}_{00}$ mode is 0.292 $\upmu$m$^2$.}

The mode profiles have been simulated both with the MPB simulation suite and with a numerical eigensolver (Lumerical MODE).
In Figures \ref{fig:WGscheme}-c), \ref{fig:WGscheme}-d) and \ref{fig:WGscheme}-e) we show the profiles of the TE and TM modes calculated with the eigensolver, under the EHM approximation for the lateral arms, \textcolor{red}{(that is, simulating a cross-section of
the structure shown in Figure 1-b)). The fundamental mode effective area is 0.348 $\mu$m$^2$. These modes agree very well with the ones obtained from MPB (not shown)}.
\textcolor{red}{Moreover,} although not perfectly, the dispersion of the three modes calculated from the eigensolver (the dots in Figure \ref{fig:Bands300}) is in agreement with the band structure calculated by MPB, apart from a relative shift of the effective refractive index; at 1550 nm the effective index of the fundamental TE mode given by the eigensolver ($n_\text{TE}$ = 1.967) is slightly higher than the one obtained from the MPB band structure (1.907).

In order to estimate the lateral confinement we calculated the effective light-lines in the horizontal direction for the TE and TM modes, which are reported as the dashed lines of Figure \ref{fig:Bands300}.
The two lines are calculated from an effective index approach:
the lateral confinement of the waveguide-arms system has been modeled by an infinite symmetric slab parallel to the $y$-$z$ plane, surrounded by a cladding material, and whose thickness equals the waveguide's width (650 nm).
The effective light-lines of the original system are then equal to the light-lines of this new slab.
The refractive index of the cladding $n_\text{TE}(\omega)$ ($n_\text{TM}(\omega)$) is the only one of importance to determine the light-line, and it is set equal to the effective index of the fundamental TE (TM) slab mode of the original 300 nm thick layer made of the EHM.
Then, the effective light-lines are described in term of the cladding index by the equations
\begin{equation}
    k_\text{TE}(\omega) = \frac{\omega}{c} n_\text{TE}(\omega),\quad k_\text{TM}(\omega) = \frac{\omega}{c} n_\text{TM}(\omega)
\end{equation}

As seen in Figure \ref{fig:Bands300}, the dispersion of the $\text{TE}_\text{01}$ mode obtained from Lumerical lies very close to the effective TE light-line, thus further confirming that the mode is only loosely bound.
Finally, tridimensional FDTD simulations confirm that indeed the eigensolver modes are guided without scattering by the full subwavelength waveguide.

We now consider sources of losses in the subwavelength waveguide other than the intrinsic material losses,
\textcolor{red}{since these may depend strongly on the fabrication process and are independent on the platform design.}
While the fundamental TE mode is found well below the light-line, one expected source of losses is given by the coupling between the fundamental waveguide mode and the modes confined in the remaining SiC layer, past the suspending arms.
Indeed, these losses vanish completely only if the width of the lateral arms $v$ is infinitely large.
\textcolor{red}{However, we find that a 2 $\upmu$m width of the arms is sufficient to suppress at negligible levels the outcoupling of the guided TE-mode. This width should also ensure the necessary mechanical strength to suspend the waveguide, since similar structures in silicon \cite{penades2014suspended} have been successfully demonstrated. The mode confinement is verified} using the same effective-index approach used to calculate the lateral effective light-lines, by adding two additional layers beyond the cladding, which is now 2 $\upmu$m thick in the $x$ direction to both the sides of the central slab.
The losses resulting from eigenmode simulations with perfectly matching layer boundary conditions are found lower than $10^{-4}$ dB/cm for $v = 2$ $\upmu$m.


Given the above results, assuming a lossless material, we expect that the main limitations of this kind of structure are given by surface roughness and disorder.
Surface roughness couples light from the guided modes to radiative modes \cite{payne1994theoretical, grillot2004size}. 
With respect to traditional ridge waveguides, we expect this effect to be slightly higher due to the presence of the additional material interfaces corresponding to the arms.
Still, if needed, the effect of roughness can be counteracted by \textcolor{red}{reducing the presence of light at the interfaces, for instance by}
increasing the width of the central branch ($w$), which increases confinement at the expense of the introduction of additional guided modes.
Disorder in the periodicity or in the position of the lateral arms also increases losses and has to be kept to low enough values. As shown in ref.~\cite{ortega2017disorder}, where these effects are studied on a similar structure to the one considered here, the jitter in the position and dimension of periodic structures should not exceed 5~nm to keep losses to a reasonable level.
Choosing a working point far below the bandgap can decrease the effect of disorder.

We now consider the nonlinear optical properties of the system.
In particular, we estimate the nonlinear waveguide parameter $\gamma$ for the nominal waveguide design.
For uniform waveguides, $\gamma$ can be defined in terms of the nonlinear Kerr index $n_2$ and of the Poynting vector $\vec{P}$ \cite{foster2004optimal}, according to
\begin{equation}
\gamma = \frac{k_0 \int_\Sigma n_2 P_z(x,y)^2\,dx\,dy}{\abs{\int P_z(x,y)\,dx\,dy}^2},
\label{eqn:uniformgamma}
\end{equation}
where $k_0$ is the vacuum wavevector, $P_z$ is the component of $\vec{P}$ normal to the integration surface and where the top integral is performed on the cross-section of the waveguide $\Sigma$, that is, where $n_2$ is non-vanishing.
Since in our case the field changes along the propagation, we average the nonlinear waveguide parameter along a single periodic cell, following the approach described in ref. \cite{sato2015rigorous}:
\begin{multline}
\gamma = \aver{\gamma(z)} = \frac{1}{a_0} \int_z^{z+a_0} \gamma(z)\,dz =\\ =\frac{k_0}{a_0}\int_z^{z+a_0}\frac{\int n_2(x,y,z) P_z(x,y,z)^2 \,dx\,dy}{\abs{\int P_z(x,y,z) \,dx\,dy}^2}\,dz
\label{eqn:gamma}
\end{multline}
where $n_2(x,y,z)$ is assumed equal to $5.31 \cdot 10^{-19}\ \text{m}^2/\text{W}$ \cite{martini2018four} where $(x,y,z)$ is found within the SiC structure and zero otherwise.
Using the Poynting field calculated from MPB, we find $\gamma = 7.346\ \text{W}^{-1} \text{m}^{-1}$.
The nonlinear waveguide parameter is calculated also using Lumerical and the EHM approximation, obtaining $\gamma = 6.182\ \text{W}^{-1} \text{m}^{-1}$; in this case, the presence of the lateral arms is taken into account by assuming that the nonlinear index of the homogeneous medium is the average of the ones of SiC and air (i.e. equal to $n_2/2 = 2.655 \cdot 10^{-19}\ \text{m}^2/\text{W}$).
By comparison, in ref. \cite{martini2018four} the nonlinear waveguide parameter of slightly lesser confining SiC waveguides was measured to be $\gamma = 3.86 \pm 0.03\ \text{W}^{-1} \text{m}^{-1}$, while the one of typical silicon nitride waveguides is close to 2 $\text{W}^{-1} \text{m}^{-1}$ \cite{tan2010group}.

\section{Tolerance}
\label{sec:tolerance}

\begin{figure}
    \centering
    \includegraphics[width=0.48\textwidth]{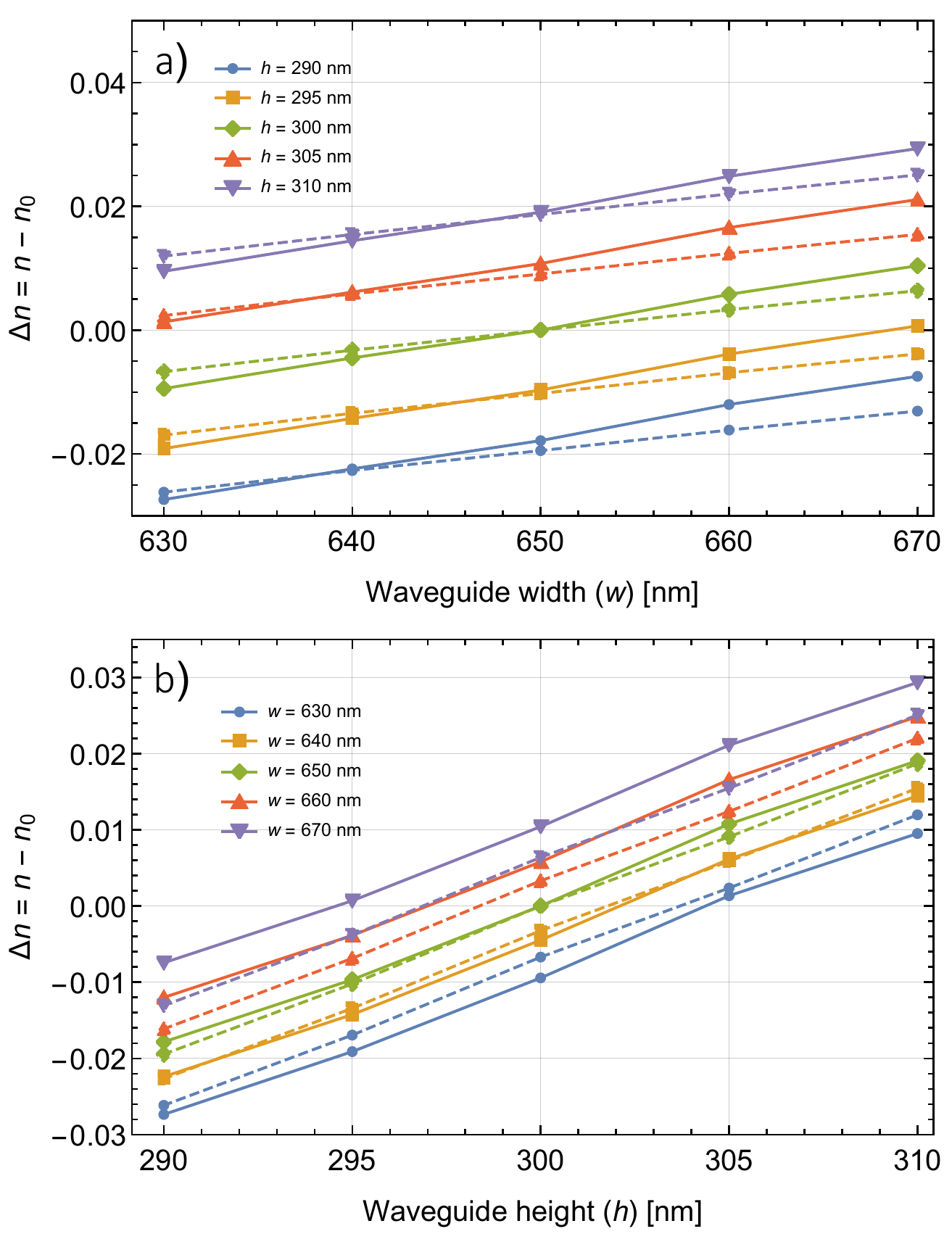}
    \caption{Simulations of the effective index $n_\text{TE}$ of the waveguide's fundamental $\text{TE}_{00}$ mode as a function of the cross-section of the central waveguide.  a) Variation of $n_\text{TE}$ in terms of the waveguide width ($w$). b) Variation of $n_\text{TE}$ in terms of the waveguide height ($h$). Continuous lines: MPB - Dashed lines: Lumerical.}
    \label{fig:tolerance}
\end{figure}

In order to determine the tolerance of the design to fabrication, we simulate the subwavelength waveguide varying its geometry; in particular, we consider variations in the cross-section and in the filling factor of the lateral arms, and we monitored the change in the effective index of the fundamental TE mode, $\Delta n = n - n_0$, where $n_0$ is the effective index of the nominal waveguide.
The simulations are performed using both the eigensolver and the MPB software suite, and the results for the cross-section variation are reported in Figure \ref{fig:tolerance}.
We find that the results obtained under the EHM approximation (dashed lines) are in good agreement with the ones obtained from MPB (continuous lines).
From the linear fit of the data obtained from MPB, we find that the sensitivities of the waveguide effective index on the width and height are respectively $\sigma_w = \Delta n/ \Delta w = 5.00 \cdot 10^{-4}\ \text{nm}^{-1}$ and $\sigma_h = \Delta n/ \Delta h = 1.88 \cdot 10^{-3}\ \text{nm}^{-1}$, while from Lumerical we find $\sigma_w = 3.27 \cdot 10^{-4}\ \text{nm}^{-1}$ and $\sigma_h = 1.91 \cdot 10^{-3}\ \text{nm}^{-1}$.
Similarly, we perform simulations varying the filling factor of the lateral arms, obtaining $\sigma_u = \Delta n/\Delta u = 9.69 \cdot 10^{-4}\ \text{nm}^{-1}$ with MPB and $\sigma_u = 1.25 \cdot 10^{-3}\ \text{nm}^{-1}$ with Lumerical.

The small overall variation of the waveguide's effective index in both cases demonstrates that the subwavelength waveguide design is very tolerant to variations of its geometrical parameters.
Moreover, these results justify the use of the homogeneous medium approximation to simplify the design procedure of other photonic devices in this platform, once the difference in the effective index given by the two simulation methods is taken into account.


\section{Photonic components}
\label{sec:devices}

A whole range of additional structures can be easily realized under the EHM approximation for the lateral arms region.
We can apply standard photonic design and simulation tools to obtain, for example, tapers and bends.
\textcolor{red}{Bending losses are obtained from 2-D eigensolver simulations under the EHM approximation, finding a value of about 0.2 dB/cm for 20 $\upmu$m bending radius.}
In order to avoid losses introduced by periodicity mismatch, the arms in bent sections have to maintain their mutual spacing as close as possible to the nominal value of the straight waveguide.
\textcolor{red}{We also simulate simple adiabatic tapers of the waveguide, finding that a lenght of 100 $\mu$m is sufficient to decrease the waveguide width from 12 $\mu$m to 650 nm with 96\% transmission.}
Again, simulations of the structures with FDTD methods confirm the expected behaviour of the devices.

\begin{figure}
    \centering
    \includegraphics[width=0.48\textwidth]{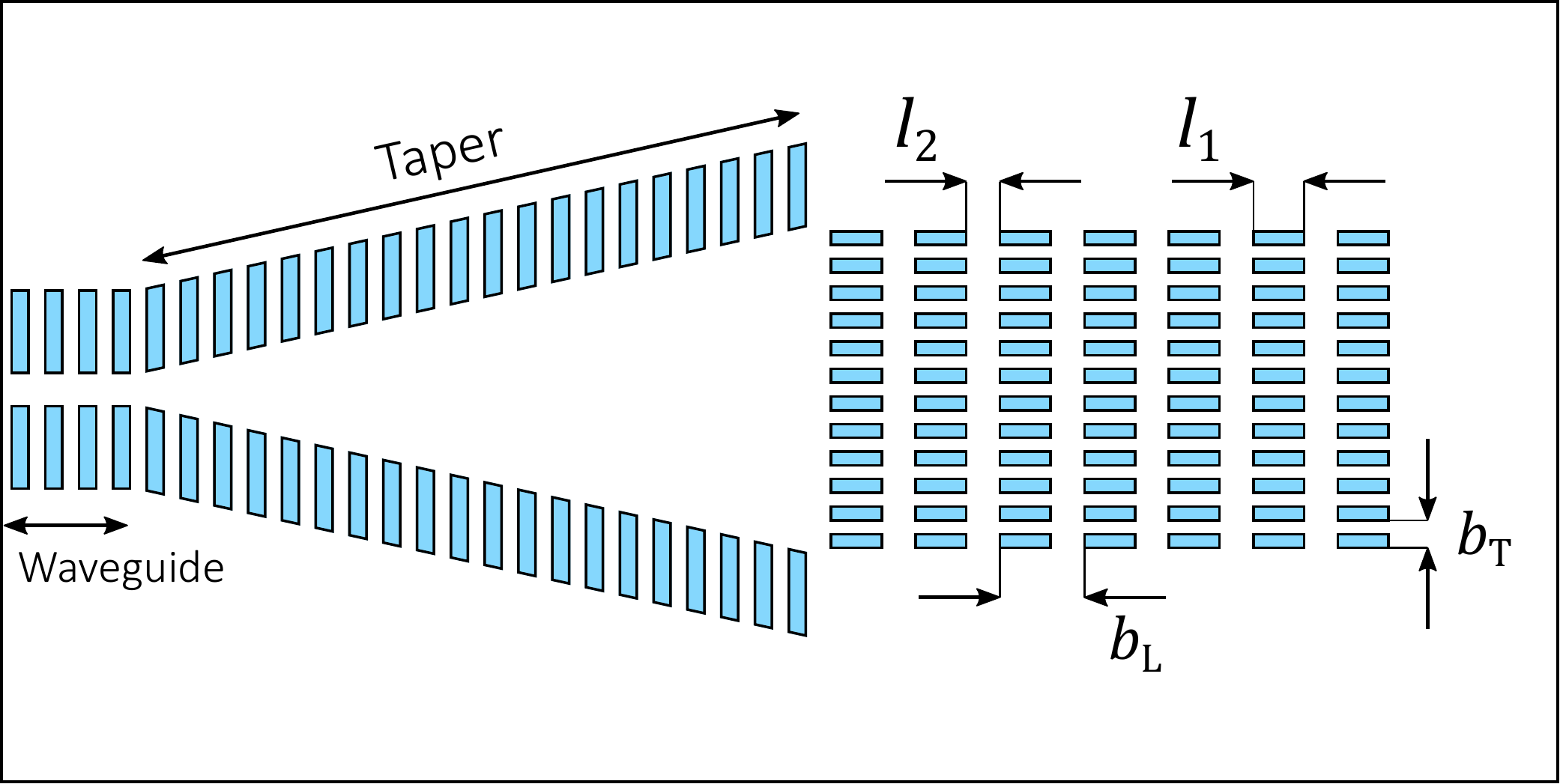}
    \caption{Schematic representation of the grating coupler geometry (top view, not to scale): blue regions represent holes to be etched in the SiC film.}
    \label{fig:Structures}
\end{figure}

Efficient coupling of light into the suspended waveguide can be achieved using grating couplers.
On this matter, different designs that exploit subwavelength structures have been proposed \cite{halir2009waveguide,cheng2012broadband, halir2010continuously,zhou2018fully,sanchez2018mid,sanchez2019design}.
Briefly, the subwavelength arms, in this case, are used to define the effective index for the subwavelength grooves of the grating coupler and thus are oriented along the propagation direction, as shown schematically in Figure \ref{fig:Structures}.
Table \ref{tab:gratingcoupler} lists the dimensions for a uniform grating coupler with simulated -3.8 dB maximum coupling efficiency, designed for TE radiation incoming at an $8\degree$ angle to normal incidence.

For the subwavelength grooves we choose a (transversal) periodicity $b_T$ and filling factor $f_\text{grat,T}$ of 300~nm and 0.5, respectively.
\textcolor{red}{The final design is obtained by first estimating the effective indices of the high and low index regions of the grating, and then by performing 3D FDTD optimization, following the design method described in ref. \cite{halir2009waveguide}.
The index estimates are used to initialize the FDTD optimization.}
\textcolor{red}{The effective index of the groove, $n_e'$, is estimated with numerical simulations of the fundamental slab mode sustained by the groove cross-section:
using Lumerical and exploting periodic boundary conditions in the transverse direction we simulate a single cell of the groove, obtaining $n'_e  = 1.110$ for $f_\text{grat,T} = 0.5$.}
\textcolor{red}{Since $n_e'$ is very close to 1, the subwavelength grooves could be replaced entirely with air, resulting in a simpler design with very similar dimensions, however suspending large sections of SiC may result in the deformation of the suspended geometries due to stress release.
The cross-section of the high index region is simply a SiC slab suspended in air, thus the associated effective index is the one of its fundamental TE mode, equal to $n''_e = 2.134$.}


\begin{table}[h]
    \centering
    \begin{tabular}{r c l}
    \hline
    \text{Film thickness} & & 300 nm\\
    \hline
    \text{Longit. period ($b_L$)} & & 1230 nm\ *\\
    \text{Longit. FF ($f_\text{grat,L}$)} & & 32.4\% *\\
    \text{Number of periods} & & 13\\
    \hline
    \text{Transv. period ($b_T$)} & & 300 nm \\
    \text{Transv. FF ($f_\text{grat,T}$)} & & 50\% \\
    \text{Transv. length} & & 12 $\mu$m\\
    \hline
    \text{Max. transmission:} & & 41.8\% (-3.8 dB)\\
    \text{1 dB bandwidth:}& & 75 nm\\
    \hline
    \end{tabular}
    \caption{Proposed dimensions and properties for a TE SiC subwavelength grating coupler operating around 1550 nm. The index of SiC is assumed to be 2.6. The values marked with * are obtained by numerical optimization.}
    \label{tab:gratingcoupler}
\end{table}

\textcolor{red}{The estimates for the longitudinal parameters of the design follow from the simplest analytic description of the uniform grating coupler:}
\begin{equation}
   l_1 = \frac{\lambda_0}{2 (n_1 - n_c \sin{\alpha})},\quad
l_2 = \frac{\lambda_0}{2 (n_2 - n_c \sin{\alpha})},
\end{equation}
\textcolor{red}{where $l_1$ and $l_2$ are the longitudinal dimensions of the low- and high-index sections of the grating (so that $b_L = l_1+l_2$ is the grating period and $f_\text{grat,L} = l_2/(l_1+l_2)$ is the grating \emph{longitudinal} filling factor), $\lambda_0$ is the vacuum wavelength of light, $\alpha$ is the angle to normal incidence, $n_1$ and $n_2$ are the low and high effective indexes of the light travelling in the grating, and $n_c$ is the index of the surrounding material. In our case $n_1$ and $n_2$ are identified with $n'_e$ and $n''_e$ and thus are equal to 1.110 and 2.134, while $n_c$ is the index of air (1.0). This gives $b_L = 1187$~nm and $l_2/(l_1+l_2) = f_\text{grat,L} = 32.7$\%.}

\begin{figure}
    \centering
    \includegraphics[width=0.48\textwidth]{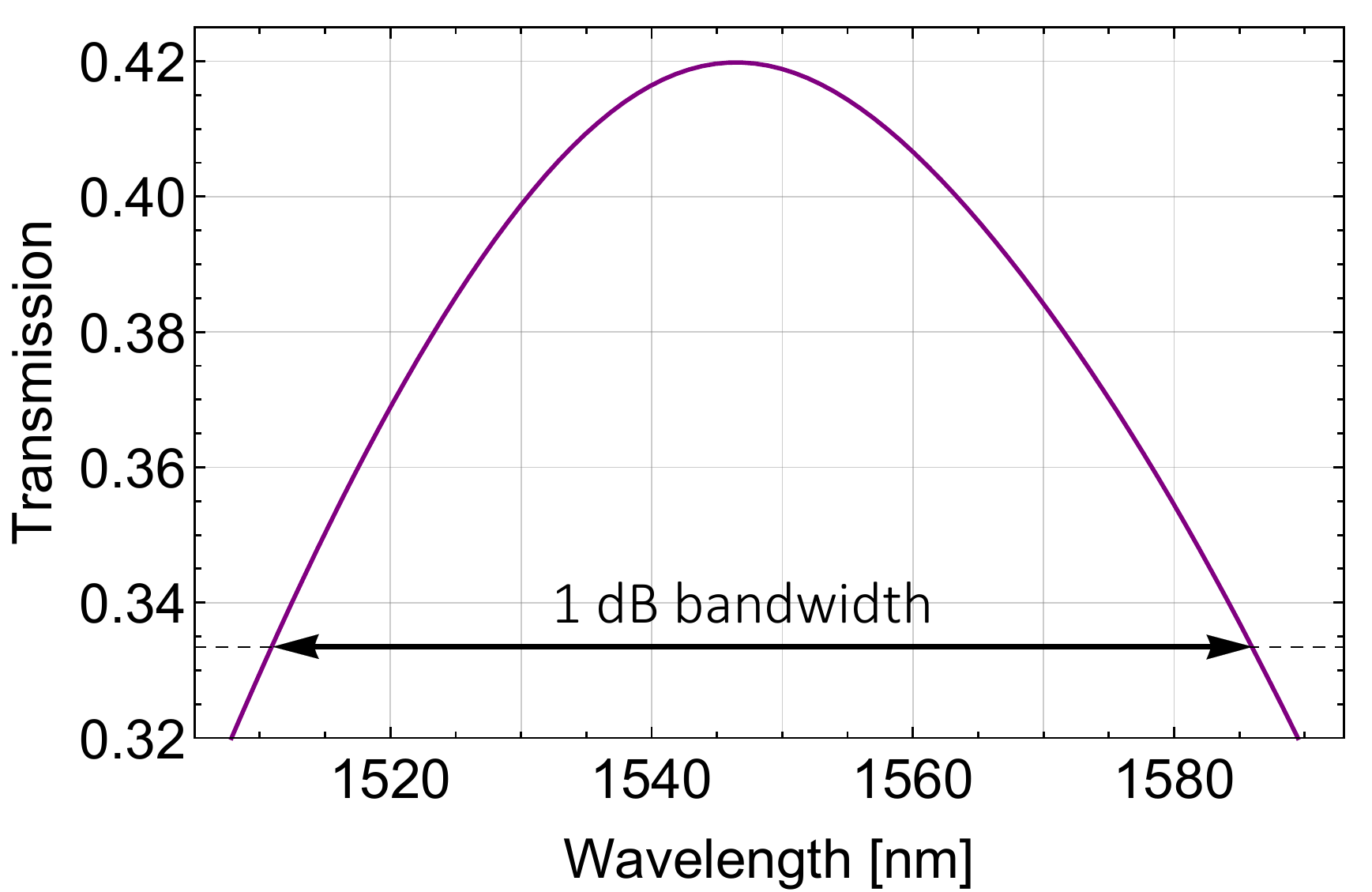}
    \caption{Simulated transmission of the proposed subwavelength grating coupler.}
    \label{fig:gratingtransmission}
\end{figure}

\textcolor{red}{Finally}, the values marked with~* in Table \ref{tab:gratingcoupler} are obtained by numerical 3D FDTD optimization (having maximum transmission at 1550~nm as target and exploiting periodic boundary conditions in the transverse direction) \textcolor{red}{and they are close to the values obtained above}.
Figure~\ref{fig:gratingtransmission} shows the transmission of the grating as a function of the wavelength, obtained from 3D FDTD simulations; The 1~dB bandwidth is 75~nm large, ranging between 1511~nm and 1586~nm.
At the expense of the bandwidth, apodised designs can be employed to increase the maximum coupling efficiency.\\

\section{Slow-light}
\label{sec:slowlight}

\begin{figure}
    \centering
    \includegraphics[width=0.48\textwidth]{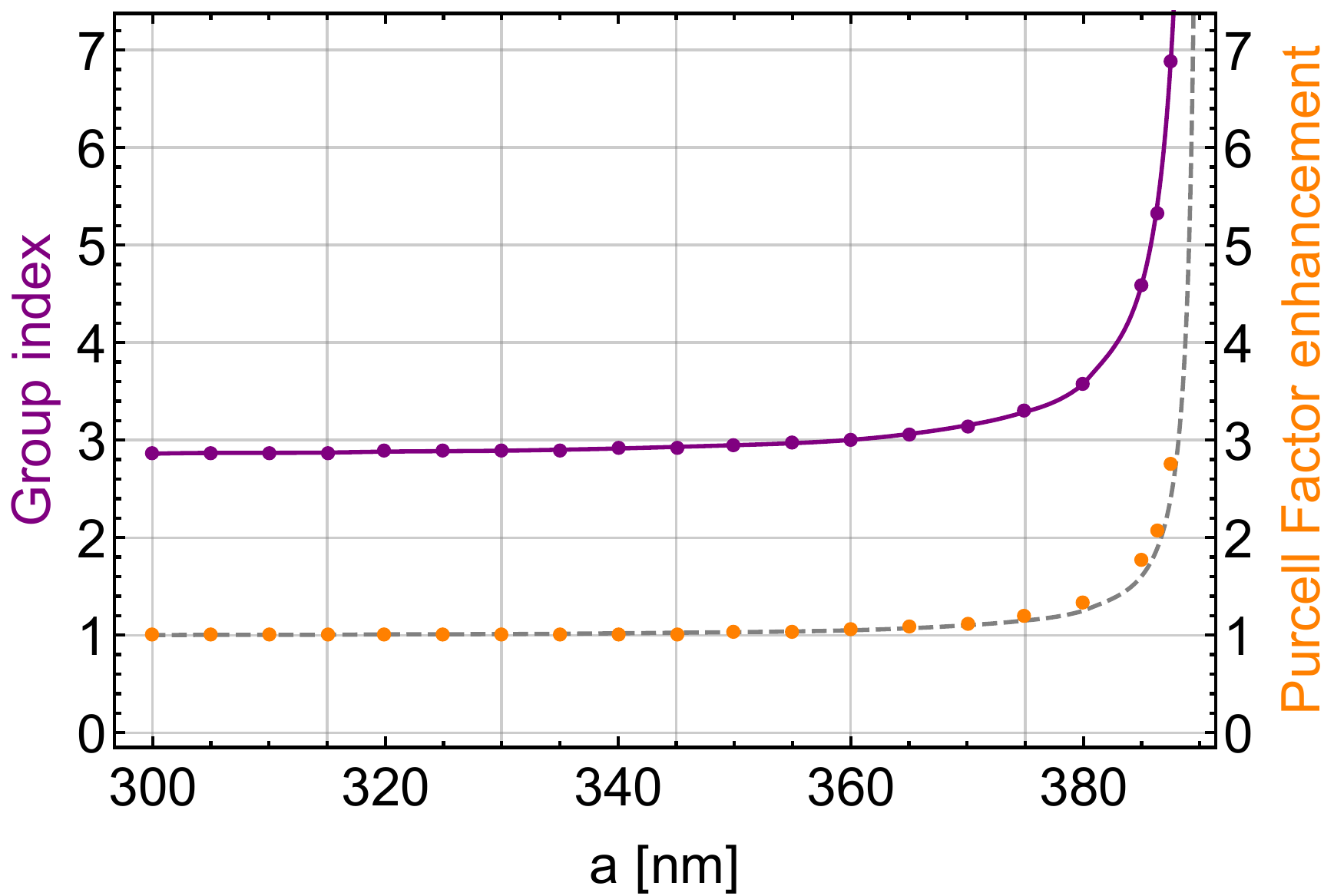}
    \caption{Group index and Purcell Factor enhancement as a function of the periodicity of the waveguide, obtained from MPB simulations. The dashed gray line is the enhancement of the group index alone, highlighting the main contribution to the Purcell Factor enhancement.}
    \label{fig:GroupIndex}
\end{figure}

In one-dimensional periodic structures, the modal dispersion near the photonic bandgap becomes flatter, corresponding to a lower group velocity of light (so called ``slow-light" regime \cite{krauss2007slow}).
Exploiting slow-light increases the interaction between radiation and matter.
For instance, the fraction of the photons emitted to a guided mode by a dipole localized in a waveguide (also known as $\beta$ factor) is inversely proportional to the group index and can reach values very close to unity \cite{rao2007single,arcari2014near}.
In our platform, the slow-light regime can be reached naturally by increasing the periodicity of the subwavelength waveguide to move down the bandgap close to the working frequency.
As it is shown in Fig. \ref{fig:GroupIndex}, the group index is more than doubled when the waveguide periodicity approaches 390~nm ($f_\text{wg}$ still equal to 0.5).

Following the approach of \cite{rao2007single}, we define an effective mode volume $V_\text{eff}$ and Purcell Factor (PF) associated to the light travelling in the waveguide.
\begin{equation}
    \text{PF} = \frac{3 \pi c^3 a}{V_\text{eff} \omega_0 \epsilon^{3/2} v_g}
\end{equation}
where $\omega_0 = 2\pi c/\lambda_0$ is the frequency of light at the working point, $\epsilon^{1/2} = n_\text{SiC}$ is the refractive index of SiC and $v_g$ is the group velocity of light; the effective volume $V_\text{eff}$ is given by
\begin{equation}
    V_\text{eff}=\frac{1}{\max(\epsilon(\vec{r})\lvert\vec{e}(\vec{r})\rvert^2)}
\end{equation}
where $\vec{e}$ is the modal electric field traveling in the waveguide, $\epsilon(\vec{r})$ is the dielectric function that defines the periodic strcture and where $\vec{r}$ is allowed to vary on the periodic cell.
Figure \ref{fig:GroupIndex} also reports the enhancement of PF (i.e. $\text{PF}(a)/\text{PF}(a_0)$, where $a$ is the increased periodicity compared to the nominal periodicity $a_0$, showing that it is mainly induced by the increase of group index.



A transition region between nominal and slow-light regimes can be realized easily by changing adiabatically the periodicity of the waveguide.
Since the field profiles of the different regions are very similar, there is no need to modulate the waveguide width to spatially match the two modes, which has been shown to be a key aspect to obtain low insertion losses \cite{krauss2007slow}.
Yet, as discussed previously, the closeness of the photon energy to the photonic bandgap would make the system more sensitive to disorder and the fabrication more challenging, and will likely set the limitation of the slow-light operation. 

The achievement of modest PF can benefit the field of quantum technologies based on color centres embedded in SiC. For example, the emission rate of the silicon vacancy (SiV) center is limited by the non-radiative decay from the excited state to a metastable state \cite{nagy2019high}. Moreover, the collection efficiency in confocal microscopy setups is hampered by the high refractive index due to TIR. A moderate PF would increase the radiative rate to values sufficient to accomplish quantum non-demolition readout of the spin state.


\section{Electro-optic Modulators}
\label{sec:modulators}

Active modulation of light travelling inside SiC can be performed with electro-optic modulators that exploit the high $\chi^{(2)}$ nonlinearity of the material.
Assuming that the bottom surface of the structure is not accessible, the modulator could be realized by patterning two metallic pads to the sides of the suspending arms' region.
In order to give an estimate of the performance of the device, we model the two pads as a parallel plate capacitor with 6~$\mu$m spacing and centered on the SiC waveguide \cite{alferness1982waveguide}, so that the overlap between the driving and optical fields is equal to unity.
Since the index of refraction of the material $n = \sqrt{1+\chi^{(1)}}$ is modified by an applied electric field $E$ according to
\begin{equation}
n(E) = \sqrt{1+\chi^{(1)}+2\chi^{(2)}E} \simeq n + \chi^{(2)}E,
\end{equation}
the standard voltage-length figure of merit for a $\pi$ phase shifter is given by
\begin{equation}
L_\pi V_\pi \simeq \frac{\lambda l}{r n^3},
\end{equation}
where $l$ the distance between the capacitor plates and $r = 2 \chi^{(2)}/n^4$ is the electro-optic coefficient of the waveguide material.
Assuming $n = 2.6$, $\lambda_0 = 1550$~nm, $l = 6\ \mu$m, $\chi^{(2)} = 32.8$ pm/V, then $r = 1.43$ pm/V and $L_\pi V_\pi \simeq 36.9\ \text{V}\cdot\text{cm}$;
this performance can be improved by a factor 2 by implementing an amplitude modulator based on a Mach-Zehnder interferometer driven by pads in the ground-signal-ground configuration, reducing $V_\pi L_\pi$ down to $18.4\ \text{V}\cdot\text{cm}$.
This value is about one order of magnitude higher than the one of state of the art electro-optic modulators based on Lithium Niobate \cite{wang2018nanophotonic} ($r_{33} \simeq 30.8$ pm/V \cite{yariv2006photonics}) which use similar spacing between the pads \cite{janner2009micro}.
As for other platforms, it is conceivable to improve the performance of electro-optic modulators using resonant structures like microring resonators \cite{xu2005micrometre}.
We also confirmed that the pads induce negligible losses: a Lumerical simulation of the optical mode propagating alongside gold pads spaced by 6 $\upmu$m and placed outside the arms' region results in about 2 dB/m losses.

\section{Conclusions}
\label{sec:conclusions}

In this work we proposed a scalable photonic platform based on SiC that allows coupling of electromagnetic radiation into and out of a SiC thin film and the manipulation of the electromagnetic field in the material.
This platform, based on suspended subwavelength waveguides, is flexible enough for the realization of all the basic photonic components such as waveguides, bends, directional couplers, grating couplers and tapers.
The proposed design requires a single etch step to access the substrate and to define the geometry of the devices, simplifying the fabrication process with respect to other previous SiC suspended platforms;
despite this, the platform retains a powerful design flexibility, because the duty cycle of subwavelength sections can be different in different parts of the sample.
As explained, an increase in the periodicity allows to reach a slow-light regime, which can be used to increase the linear and nonlinear interaction of light with SiC nonlinearities or color centers therein.
For instance, this effect can be used to shorten the length of superconducting nanowires or electro-optical modulators integrated alongside the suspended waveguides.

Since SiC is very hard, compared to other sub-wavelength platforms realized in other materials such as Si and germanium, we believe that the lateral suspending structures can be very thin, hence allowing sub-wavelength regimes for shorter wavelengths than previously demonstrated.
Reaching a periodicity shorter than 280~nm would permit the propagation of 1100~nm light, which in turn enables the interaction with NIR defects in SiC.
Quantum optics applications would then become feasible.
This, together with the increase of the $\beta$ factor given by slow-light could make this platform appealing for both 3C- and 4H-SiC.
An even shorter periodicity of 200~nm would sustain the guided propagation of 785~nm radiation, the second harmonic of 1550~nm;
provided that a suitable way to obtain phase-matching between these two frequencies can be found, the strong $\chi^{(2)}$ nonlinearity of SiC would allow the efficient exploitation of second harmonic generation and stimulated/spontaneous parametric down-conversion.

Provided that the overall losses of the platform, given not only by the material but also by roughness and disorder, can be kept low enough, squeezing on an integrated, scalable platform would become a concrete and promising application.\\

\noindent \textbf{Funding}

\noindent This work has been supported by the Engineering and Physical Sciences Research Council (EPSRC) (EP/P003710/1)\\

\noindent \textbf{Acknowledgements}

\noindent  Useful discussions with Marco Liscidini are acknowledged.


\bibliography{citations}

\begin{thebibliography}{48}%
\makeatletter
\providecommand \@ifxundefined [1]{%
 \@ifx{#1\undefined}
}%
\providecommand \@ifnum [1]{%
 \ifnum #1\expandafter \@firstoftwo
 \else \expandafter \@secondoftwo
 \fi
}%
\providecommand \@ifx [1]{%
 \ifx #1\expandafter \@firstoftwo
 \else \expandafter \@secondoftwo
 \fi
}%
\providecommand \natexlab [1]{#1}%
\providecommand \enquote  [1]{``#1''}%
\providecommand \bibnamefont  [1]{#1}%
\providecommand \bibfnamefont [1]{#1}%
\providecommand \citenamefont [1]{#1}%
\providecommand \href@noop [0]{\@secondoftwo}%
\providecommand \href [0]{\begingroup \@sanitize@url \@href}%
\providecommand \@href[1]{\@@startlink{#1}\@@href}%
\providecommand \@@href[1]{\endgroup#1\@@endlink}%
\providecommand \@sanitize@url [0]{\catcode `\\12\catcode `\$12\catcode
  `\&12\catcode `\#12\catcode `\^12\catcode `\_12\catcode `\%12\relax}%
\providecommand \@@startlink[1]{}%
\providecommand \@@endlink[0]{}%
\providecommand \url  [0]{\begingroup\@sanitize@url \@url }%
\providecommand \@url [1]{\endgroup\@href {#1}{\urlprefix }}%
\providecommand \urlprefix  [0]{URL }%
\providecommand \Eprint [0]{\href }%
\providecommand \doibase [0]{https://doi.org/}%
\providecommand \selectlanguage [0]{\@gobble}%
\providecommand \bibinfo  [0]{\@secondoftwo}%
\providecommand \bibfield  [0]{\@secondoftwo}%
\providecommand \translation [1]{[#1]}%
\providecommand \BibitemOpen [0]{}%
\providecommand \bibitemStop [0]{}%
\providecommand \bibitemNoStop [0]{.\EOS\space}%
\providecommand \EOS [0]{\spacefactor3000\relax}%
\providecommand \BibitemShut  [1]{\csname bibitem#1\endcsname}%
\let\auto@bib@innerbib\@empty
\bibitem [{\citenamefont {Koehl}\ \emph {et~al.}(2011)\citenamefont {Koehl},
  \citenamefont {Buckley}, \citenamefont {Heremans}, \citenamefont {Calusine},\
  and\ \citenamefont {Awschalom}}]{koehl2011room}%
  \BibitemOpen
  \bibfield  {author} {\bibinfo {author} {\bibfnamefont {W.~F.}\ \bibnamefont
  {Koehl}}, \bibinfo {author} {\bibfnamefont {B.~B.}\ \bibnamefont {Buckley}},
  \bibinfo {author} {\bibfnamefont {F.~J.}\ \bibnamefont {Heremans}}, \bibinfo
  {author} {\bibfnamefont {G.}~\bibnamefont {Calusine}},\ and\ \bibinfo
  {author} {\bibfnamefont {D.~D.}\ \bibnamefont {Awschalom}},\ }\bibfield
  {title} {\bibinfo {title} {Room temperature coherent control of defect spin
  qubits in silicon carbide},\ }\href@noop {} {\bibfield  {journal} {\bibinfo
  {journal} {Nature}\ }\textbf {\bibinfo {volume} {479}},\ \bibinfo {pages}
  {84} (\bibinfo {year} {2011})}\BibitemShut {NoStop}%
\bibitem [{\citenamefont {Falk}\ \emph {et~al.}(2013)\citenamefont {Falk},
  \citenamefont {Buckley}, \citenamefont {Calusine}, \citenamefont {Koehl},
  \citenamefont {Dobrovitski}, \citenamefont {Politi}, \citenamefont {Zorman},
  \citenamefont {Feng},\ and\ \citenamefont {Awschalom}}]{falk2013polytype}%
  \BibitemOpen
  \bibfield  {author} {\bibinfo {author} {\bibfnamefont {A.~L.}\ \bibnamefont
  {Falk}}, \bibinfo {author} {\bibfnamefont {B.~B.}\ \bibnamefont {Buckley}},
  \bibinfo {author} {\bibfnamefont {G.}~\bibnamefont {Calusine}}, \bibinfo
  {author} {\bibfnamefont {W.~F.}\ \bibnamefont {Koehl}}, \bibinfo {author}
  {\bibfnamefont {V.~V.}\ \bibnamefont {Dobrovitski}}, \bibinfo {author}
  {\bibfnamefont {A.}~\bibnamefont {Politi}}, \bibinfo {author} {\bibfnamefont
  {C.~A.}\ \bibnamefont {Zorman}}, \bibinfo {author} {\bibfnamefont {P.~X.-L.}\
  \bibnamefont {Feng}},\ and\ \bibinfo {author} {\bibfnamefont {D.~D.}\
  \bibnamefont {Awschalom}},\ }\bibfield  {title} {\bibinfo {title} {Polytype
  control of spin qubits in silicon carbide},\ }\href@noop {} {\bibfield
  {journal} {\bibinfo  {journal} {Nature Communications}\ }\textbf {\bibinfo
  {volume} {4}},\ \bibinfo {pages} {1819} (\bibinfo {year} {2013})}\BibitemShut
  {NoStop}%
\bibitem [{\citenamefont {Castelletto}\ \emph {et~al.}(2014)\citenamefont
  {Castelletto}, \citenamefont {Johnson}, \citenamefont {Iv{\'a}dy},
  \citenamefont {Stavrias}, \citenamefont {Umeda}, \citenamefont {Gali},\ and\
  \citenamefont {Ohshima}}]{castelletto2014silicon}%
  \BibitemOpen
  \bibfield  {author} {\bibinfo {author} {\bibfnamefont {S.}~\bibnamefont
  {Castelletto}}, \bibinfo {author} {\bibfnamefont {B.}~\bibnamefont
  {Johnson}}, \bibinfo {author} {\bibfnamefont {V.}~\bibnamefont {Iv{\'a}dy}},
  \bibinfo {author} {\bibfnamefont {N.}~\bibnamefont {Stavrias}}, \bibinfo
  {author} {\bibfnamefont {T.}~\bibnamefont {Umeda}}, \bibinfo {author}
  {\bibfnamefont {A.}~\bibnamefont {Gali}},\ and\ \bibinfo {author}
  {\bibfnamefont {T.}~\bibnamefont {Ohshima}},\ }\bibfield  {title} {\bibinfo
  {title} {A silicon carbide room-temperature single-photon source},\
  }\href@noop {} {\bibfield  {journal} {\bibinfo  {journal} {Nature Materials}\
  }\textbf {\bibinfo {volume} {13}},\ \bibinfo {pages} {151} (\bibinfo {year}
  {2014})}\BibitemShut {NoStop}%
\bibitem [{\citenamefont {Widmann}\ \emph {et~al.}(2015)\citenamefont
  {Widmann}, \citenamefont {Lee}, \citenamefont {Rendler}, \citenamefont {Son},
  \citenamefont {Fedder}, \citenamefont {Paik}, \citenamefont {Yang},
  \citenamefont {Zhao}, \citenamefont {Yang}, \citenamefont {Booker} \emph
  {et~al.}}]{widmann2015coherent}%
  \BibitemOpen
  \bibfield  {author} {\bibinfo {author} {\bibfnamefont {M.}~\bibnamefont
  {Widmann}}, \bibinfo {author} {\bibfnamefont {S.-Y.}\ \bibnamefont {Lee}},
  \bibinfo {author} {\bibfnamefont {T.}~\bibnamefont {Rendler}}, \bibinfo
  {author} {\bibfnamefont {N.~T.}\ \bibnamefont {Son}}, \bibinfo {author}
  {\bibfnamefont {H.}~\bibnamefont {Fedder}}, \bibinfo {author} {\bibfnamefont
  {S.}~\bibnamefont {Paik}}, \bibinfo {author} {\bibfnamefont {L.-P.}\
  \bibnamefont {Yang}}, \bibinfo {author} {\bibfnamefont {N.}~\bibnamefont
  {Zhao}}, \bibinfo {author} {\bibfnamefont {S.}~\bibnamefont {Yang}}, \bibinfo
  {author} {\bibfnamefont {I.}~\bibnamefont {Booker}}, \emph {et~al.},\
  }\bibfield  {title} {\bibinfo {title} {Coherent control of single spins in
  silicon carbide at room temperature},\ }\href@noop {} {\bibfield  {journal}
  {\bibinfo  {journal} {Nature Materials}\ }\textbf {\bibinfo {volume} {14}},\
  \bibinfo {pages} {164} (\bibinfo {year} {2015})}\BibitemShut {NoStop}%
\bibitem [{\citenamefont {Christle}\ \emph {et~al.}(2015)\citenamefont
  {Christle}, \citenamefont {Falk}, \citenamefont {Andrich}, \citenamefont
  {Klimov}, \citenamefont {Hassan}, \citenamefont {Son}, \citenamefont
  {Janz{\'e}n}, \citenamefont {Ohshima},\ and\ \citenamefont
  {Awschalom}}]{christle2015isolated}%
  \BibitemOpen
  \bibfield  {author} {\bibinfo {author} {\bibfnamefont {D.~J.}\ \bibnamefont
  {Christle}}, \bibinfo {author} {\bibfnamefont {A.~L.}\ \bibnamefont {Falk}},
  \bibinfo {author} {\bibfnamefont {P.}~\bibnamefont {Andrich}}, \bibinfo
  {author} {\bibfnamefont {P.~V.}\ \bibnamefont {Klimov}}, \bibinfo {author}
  {\bibfnamefont {J.~U.}\ \bibnamefont {Hassan}}, \bibinfo {author}
  {\bibfnamefont {N.~T.}\ \bibnamefont {Son}}, \bibinfo {author} {\bibfnamefont
  {E.}~\bibnamefont {Janz{\'e}n}}, \bibinfo {author} {\bibfnamefont
  {T.}~\bibnamefont {Ohshima}},\ and\ \bibinfo {author} {\bibfnamefont {D.~D.}\
  \bibnamefont {Awschalom}},\ }\bibfield  {title} {\bibinfo {title} {Isolated
  electron spins in silicon carbide with millisecond coherence times},\
  }\href@noop {} {\bibfield  {journal} {\bibinfo  {journal} {Nature Materials}\
  }\textbf {\bibinfo {volume} {14}},\ \bibinfo {pages} {160} (\bibinfo {year}
  {2015})}\BibitemShut {NoStop}%
\bibitem [{\citenamefont {Christle}\ \emph {et~al.}(2017)\citenamefont
  {Christle}, \citenamefont {Klimov}, \citenamefont {Charles}, \citenamefont
  {Sz{\'a}sz}, \citenamefont {Iv{\'a}dy}, \citenamefont {Jokubavicius},
  \citenamefont {Hassan}, \citenamefont {Syv{\"a}j{\"a}rvi}, \citenamefont
  {Koehl}, \citenamefont {Ohshima} \emph {et~al.}}]{christle2017isolated}%
  \BibitemOpen
  \bibfield  {author} {\bibinfo {author} {\bibfnamefont {D.~J.}\ \bibnamefont
  {Christle}}, \bibinfo {author} {\bibfnamefont {P.~V.}\ \bibnamefont
  {Klimov}}, \bibinfo {author} {\bibfnamefont {F.}~\bibnamefont {Charles}},
  \bibinfo {author} {\bibfnamefont {K.}~\bibnamefont {Sz{\'a}sz}}, \bibinfo
  {author} {\bibfnamefont {V.}~\bibnamefont {Iv{\'a}dy}}, \bibinfo {author}
  {\bibfnamefont {V.}~\bibnamefont {Jokubavicius}}, \bibinfo {author}
  {\bibfnamefont {J.~U.}\ \bibnamefont {Hassan}}, \bibinfo {author}
  {\bibfnamefont {M.}~\bibnamefont {Syv{\"a}j{\"a}rvi}}, \bibinfo {author}
  {\bibfnamefont {W.~F.}\ \bibnamefont {Koehl}}, \bibinfo {author}
  {\bibfnamefont {T.}~\bibnamefont {Ohshima}}, \emph {et~al.},\ }\bibfield
  {title} {\bibinfo {title} {Isolated spin qubits in {SiC} with a high-fidelity
  infrared spin-to-photon interface},\ }\href@noop {} {\bibfield  {journal}
  {\bibinfo  {journal} {Physical Review X}\ }\textbf {\bibinfo {volume} {7}},\
  \bibinfo {pages} {021046} (\bibinfo {year} {2017})}\BibitemShut {NoStop}%
\bibitem [{\citenamefont {Simin}\ \emph {et~al.}(2017)\citenamefont {Simin},
  \citenamefont {Kraus}, \citenamefont {Sperlich}, \citenamefont {Ohshima},
  \citenamefont {Astakhov},\ and\ \citenamefont {Dyakonov}}]{simin2017locking}%
  \BibitemOpen
  \bibfield  {author} {\bibinfo {author} {\bibfnamefont {D.}~\bibnamefont
  {Simin}}, \bibinfo {author} {\bibfnamefont {H.}~\bibnamefont {Kraus}},
  \bibinfo {author} {\bibfnamefont {A.}~\bibnamefont {Sperlich}}, \bibinfo
  {author} {\bibfnamefont {T.}~\bibnamefont {Ohshima}}, \bibinfo {author}
  {\bibfnamefont {G.}~\bibnamefont {Astakhov}},\ and\ \bibinfo {author}
  {\bibfnamefont {V.}~\bibnamefont {Dyakonov}},\ }\bibfield  {title} {\bibinfo
  {title} {Locking of electron spin coherence above 20 ms in natural silicon
  carbide},\ }\href@noop {} {\bibfield  {journal} {\bibinfo  {journal}
  {Physical Review B}\ }\textbf {\bibinfo {volume} {95}},\ \bibinfo {pages}
  {161201} (\bibinfo {year} {2017})}\BibitemShut {NoStop}%
\bibitem [{\citenamefont {Tang}\ \emph {et~al.}(1991)\citenamefont {Tang},
  \citenamefont {Irvine}, \citenamefont {Zhang},\ and\ \citenamefont
  {Spencer}}]{tang1991linear}%
  \BibitemOpen
  \bibfield  {author} {\bibinfo {author} {\bibfnamefont {X.}~\bibnamefont
  {Tang}}, \bibinfo {author} {\bibfnamefont {K.~G.}\ \bibnamefont {Irvine}},
  \bibinfo {author} {\bibfnamefont {D.}~\bibnamefont {Zhang}},\ and\ \bibinfo
  {author} {\bibfnamefont {M.~G.}\ \bibnamefont {Spencer}},\ }\bibfield
  {title} {\bibinfo {title} {Linear electro-optic effect in cubic silicon
  carbide},\ }\href@noop {} {\bibfield  {journal} {\bibinfo  {journal} {Applied
  Physics Letters}\ }\textbf {\bibinfo {volume} {59}},\ \bibinfo {pages} {1938}
  (\bibinfo {year} {1991})}\BibitemShut {NoStop}%
\bibitem [{\citenamefont {Wu}\ and\ \citenamefont {Guo}(2008)}]{wu2008second}%
  \BibitemOpen
  \bibfield  {author} {\bibinfo {author} {\bibfnamefont {I.}~\bibnamefont
  {Wu}}\ and\ \bibinfo {author} {\bibfnamefont {G.}~\bibnamefont {Guo}},\
  }\bibfield  {title} {\bibinfo {title} {Second-harmonic generation and linear
  electro-optical coefficients of {SiC} polytypes and nanotubes},\ }\href@noop
  {} {\bibfield  {journal} {\bibinfo  {journal} {Physical Review B}\ }\textbf
  {\bibinfo {volume} {78}},\ \bibinfo {pages} {035447} (\bibinfo {year}
  {2008})}\BibitemShut {NoStop}%
\bibitem [{\citenamefont {Madelung}(1982)}]{madelung1982physics}%
  \BibitemOpen
  \bibfield  {author} {\bibinfo {author} {\bibfnamefont {O.}~\bibnamefont
  {Madelung}},\ }\bibfield  {title} {\bibinfo {title} {Physics of group {IV}
  elements and {III-V} compounds},\ }\href@noop {} {\bibfield  {journal}
  {\bibinfo  {journal} {Landolt-Bornstein, Numerical Data and Functional
  Relationships in Science and Technology, New Series}\ }\textbf {\bibinfo
  {volume} {17}} (\bibinfo {year} {1982})}\BibitemShut {NoStop}%
\bibitem [{\citenamefont {Jackson}\ \emph {et~al.}(2005)\citenamefont
  {Jackson}, \citenamefont {Dunning}, \citenamefont {Zorman}, \citenamefont
  {Mehregany},\ and\ \citenamefont {Sharpe}}]{jackson2005mechanical}%
  \BibitemOpen
  \bibfield  {author} {\bibinfo {author} {\bibfnamefont {K.~M.}\ \bibnamefont
  {Jackson}}, \bibinfo {author} {\bibfnamefont {J.}~\bibnamefont {Dunning}},
  \bibinfo {author} {\bibfnamefont {C.~A.}\ \bibnamefont {Zorman}}, \bibinfo
  {author} {\bibfnamefont {M.}~\bibnamefont {Mehregany}},\ and\ \bibinfo
  {author} {\bibfnamefont {W.~N.}\ \bibnamefont {Sharpe}},\ }\bibfield  {title}
  {\bibinfo {title} {Mechanical properties of epitaxial {3C} silicon carbide
  thin films},\ }\href@noop {} {\bibfield  {journal} {\bibinfo  {journal}
  {Journal of microelectromechanical systems}\ }\textbf {\bibinfo {volume}
  {14}},\ \bibinfo {pages} {664} (\bibinfo {year} {2005})}\BibitemShut
  {NoStop}%
\bibitem [{\citenamefont {Chatzopoulos}\ \emph {et~al.}(2019)\citenamefont
  {Chatzopoulos}, \citenamefont {Martini}, \citenamefont {Cernansky},\ and\
  \citenamefont {Politi}}]{chatzopoulos2019high}%
  \BibitemOpen
  \bibfield  {author} {\bibinfo {author} {\bibfnamefont {I.}~\bibnamefont
  {Chatzopoulos}}, \bibinfo {author} {\bibfnamefont {F.}~\bibnamefont
  {Martini}}, \bibinfo {author} {\bibfnamefont {R.}~\bibnamefont {Cernansky}},\
  and\ \bibinfo {author} {\bibfnamefont {A.}~\bibnamefont {Politi}},\
  }\bibfield  {title} {\bibinfo {title} {{High-Q/V} photonic crystal cavities
  and {QED} analysis in {3C-SiC}},\ }\href@noop {} {\bibfield  {journal}
  {\bibinfo  {journal} {ACS Photonics}\ }\textbf {\bibinfo {volume} {6}},\
  \bibinfo {pages} {1826} (\bibinfo {year} {2019})}\BibitemShut {NoStop}%
\bibitem [{\citenamefont {Caldwell}\ \emph {et~al.}(2013)\citenamefont
  {Caldwell}, \citenamefont {Glembocki}, \citenamefont {Francescato},
  \citenamefont {Sharac}, \citenamefont {Giannini}, \citenamefont {Bezares},
  \citenamefont {Long}, \citenamefont {Owrutsky}, \citenamefont {Vurgaftman},
  \citenamefont {Tischler} \emph {et~al.}}]{caldwell2013low}%
  \BibitemOpen
  \bibfield  {author} {\bibinfo {author} {\bibfnamefont {J.~D.}\ \bibnamefont
  {Caldwell}}, \bibinfo {author} {\bibfnamefont {O.~J.}\ \bibnamefont
  {Glembocki}}, \bibinfo {author} {\bibfnamefont {Y.}~\bibnamefont
  {Francescato}}, \bibinfo {author} {\bibfnamefont {N.}~\bibnamefont {Sharac}},
  \bibinfo {author} {\bibfnamefont {V.}~\bibnamefont {Giannini}}, \bibinfo
  {author} {\bibfnamefont {F.~J.}\ \bibnamefont {Bezares}}, \bibinfo {author}
  {\bibfnamefont {J.~P.}\ \bibnamefont {Long}}, \bibinfo {author}
  {\bibfnamefont {J.~C.}\ \bibnamefont {Owrutsky}}, \bibinfo {author}
  {\bibfnamefont {I.}~\bibnamefont {Vurgaftman}}, \bibinfo {author}
  {\bibfnamefont {J.~G.}\ \bibnamefont {Tischler}}, \emph {et~al.},\ }\bibfield
   {title} {\bibinfo {title} {Low-loss, extreme subdiffraction photon
  confinement via silicon carbide localized surface phonon polariton
  resonators},\ }\href@noop {} {\bibfield  {journal} {\bibinfo  {journal} {Nano
  letters}\ }\textbf {\bibinfo {volume} {13}},\ \bibinfo {pages} {3690}
  (\bibinfo {year} {2013})}\BibitemShut {NoStop}%
\bibitem [{\citenamefont {Fan}\ \emph {et~al.}(2018)\citenamefont {Fan},
  \citenamefont {Moradinejad}, \citenamefont {Wu}, \citenamefont {Eftekhar},\
  and\ \citenamefont {Adibi}}]{fan2018high}%
  \BibitemOpen
  \bibfield  {author} {\bibinfo {author} {\bibfnamefont {T.}~\bibnamefont
  {Fan}}, \bibinfo {author} {\bibfnamefont {H.}~\bibnamefont {Moradinejad}},
  \bibinfo {author} {\bibfnamefont {X.}~\bibnamefont {Wu}}, \bibinfo {author}
  {\bibfnamefont {A.~A.}\ \bibnamefont {Eftekhar}},\ and\ \bibinfo {author}
  {\bibfnamefont {A.}~\bibnamefont {Adibi}},\ }\bibfield  {title} {\bibinfo
  {title} {{High-Q} integrated photonic microresonators on
  {3C-SiC-on-insulator} ({SiCOI}) platform},\ }\href@noop {} {\bibfield
  {journal} {\bibinfo  {journal} {Optics Express}\ }\textbf {\bibinfo {volume}
  {26}},\ \bibinfo {pages} {25814} (\bibinfo {year} {2018})}\BibitemShut
  {NoStop}%
\bibitem [{\citenamefont {Di~Cioccio}\ \emph {et~al.}(1996)\citenamefont
  {Di~Cioccio}, \citenamefont {Le~Tiec}, \citenamefont {Letertre},
  \citenamefont {Jaussaud},\ and\ \citenamefont {Bruel}}]{di1996silicon}%
  \BibitemOpen
  \bibfield  {author} {\bibinfo {author} {\bibfnamefont {L.}~\bibnamefont
  {Di~Cioccio}}, \bibinfo {author} {\bibfnamefont {Y.}~\bibnamefont {Le~Tiec}},
  \bibinfo {author} {\bibfnamefont {F.}~\bibnamefont {Letertre}}, \bibinfo
  {author} {\bibfnamefont {C.}~\bibnamefont {Jaussaud}},\ and\ \bibinfo
  {author} {\bibfnamefont {M.}~\bibnamefont {Bruel}},\ }\bibfield  {title}
  {\bibinfo {title} {Silicon carbide on insulator formation using the {Smart}
  {Cut} process},\ }\href@noop {} {\bibfield  {journal} {\bibinfo  {journal}
  {Electronics Letters}\ }\textbf {\bibinfo {volume} {32}},\ \bibinfo {pages}
  {1144} (\bibinfo {year} {1996})}\BibitemShut {NoStop}%
\bibitem [{\citenamefont {Song}\ \emph {et~al.}(ts)\citenamefont {Song},
  \citenamefont {Asano}, \citenamefont {Jeon}, \citenamefont {Kim},
  \citenamefont {Chen}, \citenamefont {Kang},\ and\ \citenamefont
  {Noda}}]{Song2019ultrahigh}%
  \BibitemOpen
  \bibfield  {author} {\bibinfo {author} {\bibfnamefont {B.-S.}\ \bibnamefont
  {Song}}, \bibinfo {author} {\bibfnamefont {T.}~\bibnamefont {Asano}},
  \bibinfo {author} {\bibfnamefont {S.}~\bibnamefont {Jeon}}, \bibinfo {author}
  {\bibfnamefont {H.}~\bibnamefont {Kim}}, \bibinfo {author} {\bibfnamefont
  {C.}~\bibnamefont {Chen}}, \bibinfo {author} {\bibfnamefont {D.~D.}\
  \bibnamefont {Kang}},\ and\ \bibinfo {author} {\bibfnamefont
  {S.}~\bibnamefont {Noda}},\ }\bibfield  {title} {\bibinfo {title}
  {{Ultrahigh-Q} photonic crystal nanocavities based on {4H} silicon carbide},\
  }\href {https://doi.org/10.1364/OPTICA.6.000991} {\bibfield  {journal}
  {\bibinfo  {journal} {Optica}\ }\textbf {\bibinfo {volume} {6}},\ \bibinfo
  {pages} {991} (\bibinfo {year} {ts})}\BibitemShut {NoStop}%
\bibitem [{\citenamefont {Lukin}\ \emph {et~al.}(2019)\citenamefont {Lukin},
  \citenamefont {Dory}, \citenamefont {Guidry}, \citenamefont {Yang},
  \citenamefont {Mishra}, \citenamefont {Trivedi}, \citenamefont {Radulaski},
  \citenamefont {Sun}, \citenamefont {Vercruysse}, \citenamefont {Ahn} \emph
  {et~al.}}]{lukin20194h}%
  \BibitemOpen
  \bibfield  {author} {\bibinfo {author} {\bibfnamefont {D.~M.}\ \bibnamefont
  {Lukin}}, \bibinfo {author} {\bibfnamefont {C.}~\bibnamefont {Dory}},
  \bibinfo {author} {\bibfnamefont {M.~A.}\ \bibnamefont {Guidry}}, \bibinfo
  {author} {\bibfnamefont {K.~Y.}\ \bibnamefont {Yang}}, \bibinfo {author}
  {\bibfnamefont {S.~D.}\ \bibnamefont {Mishra}}, \bibinfo {author}
  {\bibfnamefont {R.}~\bibnamefont {Trivedi}}, \bibinfo {author} {\bibfnamefont
  {M.}~\bibnamefont {Radulaski}}, \bibinfo {author} {\bibfnamefont
  {S.}~\bibnamefont {Sun}}, \bibinfo {author} {\bibfnamefont {D.}~\bibnamefont
  {Vercruysse}}, \bibinfo {author} {\bibfnamefont {G.~H.}\ \bibnamefont {Ahn}},
  \emph {et~al.},\ }\bibfield  {title} {\bibinfo {title}
  {{4H}-silicon-carbide-on-insulator for integrated quantum and nonlinear
  photonics},\ }\bibfield  {journal} {\bibinfo  {journal} {Nature Photonics}\
  }\href {https://doi.org/10.1038/s41566-019-0556-6}
  {10.1038/s41566-019-0556-6} (\bibinfo {year} {2019})\BibitemShut {NoStop}%
\bibitem [{\citenamefont {Bracher}\ and\ \citenamefont
  {Hu}(2015)}]{bracher2015fabrication}%
  \BibitemOpen
  \bibfield  {author} {\bibinfo {author} {\bibfnamefont {D.~O.}\ \bibnamefont
  {Bracher}}\ and\ \bibinfo {author} {\bibfnamefont {E.~L.}\ \bibnamefont
  {Hu}},\ }\bibfield  {title} {\bibinfo {title} {Fabrication of {high-Q}
  nanobeam photonic crystals in epitaxially grown {4H-SiC}},\ }\href@noop {}
  {\bibfield  {journal} {\bibinfo  {journal} {Nano Letters}\ }\textbf {\bibinfo
  {volume} {15}},\ \bibinfo {pages} {6202} (\bibinfo {year}
  {2015})}\BibitemShut {NoStop}%
\bibitem [{\citenamefont {Calusine}\ \emph {et~al.}(2014)\citenamefont
  {Calusine}, \citenamefont {Politi},\ and\ \citenamefont
  {Awschalom}}]{calusine2014silicon}%
  \BibitemOpen
  \bibfield  {author} {\bibinfo {author} {\bibfnamefont {G.}~\bibnamefont
  {Calusine}}, \bibinfo {author} {\bibfnamefont {A.}~\bibnamefont {Politi}},\
  and\ \bibinfo {author} {\bibfnamefont {D.~D.}\ \bibnamefont {Awschalom}},\
  }\bibfield  {title} {\bibinfo {title} {Silicon carbide photonic crystal
  cavities with integrated color centers},\ }\href@noop {} {\bibfield
  {journal} {\bibinfo  {journal} {Applied Physics Letters}\ }\textbf {\bibinfo
  {volume} {105}},\ \bibinfo {pages} {011123} (\bibinfo {year}
  {2014})}\BibitemShut {NoStop}%
\bibitem [{\citenamefont {Martini}\ and\ \citenamefont
  {Politi}(2017)}]{martini2017linear}%
  \BibitemOpen
  \bibfield  {author} {\bibinfo {author} {\bibfnamefont {F.}~\bibnamefont
  {Martini}}\ and\ \bibinfo {author} {\bibfnamefont {A.}~\bibnamefont
  {Politi}},\ }\bibfield  {title} {\bibinfo {title} {Linear integrated optics
  in {3C} silicon carbide},\ }\href@noop {} {\bibfield  {journal} {\bibinfo
  {journal} {Optics Express}\ }\textbf {\bibinfo {volume} {25}},\ \bibinfo
  {pages} {10735} (\bibinfo {year} {2017})}\BibitemShut {NoStop}%
\bibitem [{\citenamefont {Penad{\'e}s}\ \emph {et~al.}(2014)\citenamefont
  {Penad{\'e}s}, \citenamefont {Alonso-Ramos}, \citenamefont {Khokhar},
  \citenamefont {Nedeljkovic}, \citenamefont {Boodhoo}, \citenamefont
  {Ortega-Mo{\~n}ux}, \citenamefont {Molina-Fern{\'a}ndez}, \citenamefont
  {Cheben},\ and\ \citenamefont {Mashanovich}}]{penades2014suspended}%
  \BibitemOpen
  \bibfield  {author} {\bibinfo {author} {\bibfnamefont {J.~S.}\ \bibnamefont
  {Penad{\'e}s}}, \bibinfo {author} {\bibfnamefont {C.}~\bibnamefont
  {Alonso-Ramos}}, \bibinfo {author} {\bibfnamefont {A.}~\bibnamefont
  {Khokhar}}, \bibinfo {author} {\bibfnamefont {M.}~\bibnamefont
  {Nedeljkovic}}, \bibinfo {author} {\bibfnamefont {L.}~\bibnamefont
  {Boodhoo}}, \bibinfo {author} {\bibfnamefont {A.}~\bibnamefont
  {Ortega-Mo{\~n}ux}}, \bibinfo {author} {\bibfnamefont {I.}~\bibnamefont
  {Molina-Fern{\'a}ndez}}, \bibinfo {author} {\bibfnamefont {P.}~\bibnamefont
  {Cheben}},\ and\ \bibinfo {author} {\bibfnamefont {G.}~\bibnamefont
  {Mashanovich}},\ }\bibfield  {title} {\bibinfo {title} {Suspended soi
  waveguide with sub-wavelength grating cladding for mid-infrared},\
  }\href@noop {} {\bibfield  {journal} {\bibinfo  {journal} {Optics Letters}\
  }\textbf {\bibinfo {volume} {39}},\ \bibinfo {pages} {5661} (\bibinfo {year}
  {2014})}\BibitemShut {NoStop}%
\bibitem [{\citenamefont {Penades}\ \emph {et~al.}(2016)\citenamefont
  {Penades}, \citenamefont {{n}ux}, \citenamefont {Nedeljkovic}, \citenamefont
  {Wang\"{u}emert-P\'{e}rez}, \citenamefont {Halir}, \citenamefont {Khokhar},
  \citenamefont {Alonso-Ramos}, \citenamefont {Qu}, \citenamefont
  {Molina-Fern\'{a}ndez}, \citenamefont {Cheben},\ and\ \citenamefont
  {Mashanovich}}]{Penades16}%
  \BibitemOpen
  \bibfield  {author} {\bibinfo {author} {\bibfnamefont {J.~S.}\ \bibnamefont
  {Penades}}, \bibinfo {author} {\bibfnamefont {A.~O.-M.}\ \bibnamefont
  {{n}ux}}, \bibinfo {author} {\bibfnamefont {M.}~\bibnamefont {Nedeljkovic}},
  \bibinfo {author} {\bibfnamefont {J.~G.}\ \bibnamefont
  {Wang\"{u}emert-P\'{e}rez}}, \bibinfo {author} {\bibfnamefont
  {R.}~\bibnamefont {Halir}}, \bibinfo {author} {\bibfnamefont {A.~Z.}\
  \bibnamefont {Khokhar}}, \bibinfo {author} {\bibfnamefont {C.}~\bibnamefont
  {Alonso-Ramos}}, \bibinfo {author} {\bibfnamefont {Z.}~\bibnamefont {Qu}},
  \bibinfo {author} {\bibfnamefont {I.}~\bibnamefont {Molina-Fern\'{a}ndez}},
  \bibinfo {author} {\bibfnamefont {P.}~\bibnamefont {Cheben}},\ and\ \bibinfo
  {author} {\bibfnamefont {G.~Z.}\ \bibnamefont {Mashanovich}},\ }\bibfield
  {title} {\bibinfo {title} {Suspended silicon mid-infrared waveguide devices
  with subwavelength grating metamaterial cladding},\ }\href
  {https://doi.org/10.1364/OE.24.022908} {\bibfield  {journal} {\bibinfo
  {journal} {Optics Express}\ }\textbf {\bibinfo {volume} {24}},\ \bibinfo
  {pages} {22908} (\bibinfo {year} {2016})}\BibitemShut {NoStop}%
\bibitem [{\citenamefont {Penad{\'e}s}\ \emph {et~al.}(2018)\citenamefont
  {Penad{\'e}s}, \citenamefont {S{\'a}nchez-Postigo}, \citenamefont
  {Nedeljkovic}, \citenamefont {Ortega-Mo{\~n}ux}, \citenamefont
  {Wang{\"u}emert-P{\'e}rez}, \citenamefont {Xu}, \citenamefont {Halir},
  \citenamefont {Qu}, \citenamefont {Khokhar}, \citenamefont {Osman} \emph
  {et~al.}}]{penades2018suspended}%
  \BibitemOpen
  \bibfield  {author} {\bibinfo {author} {\bibfnamefont {J.~S.}\ \bibnamefont
  {Penad{\'e}s}}, \bibinfo {author} {\bibfnamefont {A.}~\bibnamefont
  {S{\'a}nchez-Postigo}}, \bibinfo {author} {\bibfnamefont {M.}~\bibnamefont
  {Nedeljkovic}}, \bibinfo {author} {\bibfnamefont {A.}~\bibnamefont
  {Ortega-Mo{\~n}ux}}, \bibinfo {author} {\bibfnamefont {J.}~\bibnamefont
  {Wang{\"u}emert-P{\'e}rez}}, \bibinfo {author} {\bibfnamefont
  {Y.}~\bibnamefont {Xu}}, \bibinfo {author} {\bibfnamefont {R.}~\bibnamefont
  {Halir}}, \bibinfo {author} {\bibfnamefont {Z.}~\bibnamefont {Qu}}, \bibinfo
  {author} {\bibfnamefont {A.}~\bibnamefont {Khokhar}}, \bibinfo {author}
  {\bibfnamefont {A.}~\bibnamefont {Osman}}, \emph {et~al.},\ }\bibfield
  {title} {\bibinfo {title} {Suspended silicon waveguides for long-wave
  infrared wavelengths},\ }\href@noop {} {\bibfield  {journal} {\bibinfo
  {journal} {Optics Letters}\ }\textbf {\bibinfo {volume} {43}},\ \bibinfo
  {pages} {795} (\bibinfo {year} {2018})}\BibitemShut {NoStop}%
\bibitem [{\citenamefont {Zhou}\ \emph {et~al.}(2018)\citenamefont {Zhou},
  \citenamefont {Cheng}, \citenamefont {Wu}, \citenamefont {Sun},\ and\
  \citenamefont {Tsang}}]{zhou2018fully}%
  \BibitemOpen
  \bibfield  {author} {\bibinfo {author} {\bibfnamefont {W.}~\bibnamefont
  {Zhou}}, \bibinfo {author} {\bibfnamefont {Z.}~\bibnamefont {Cheng}},
  \bibinfo {author} {\bibfnamefont {X.}~\bibnamefont {Wu}}, \bibinfo {author}
  {\bibfnamefont {X.}~\bibnamefont {Sun}},\ and\ \bibinfo {author}
  {\bibfnamefont {H.~K.}\ \bibnamefont {Tsang}},\ }\bibfield  {title} {\bibinfo
  {title} {Fully suspended slot waveguide platform},\ }\href@noop {} {\bibfield
   {journal} {\bibinfo  {journal} {Journal of Applied Physics}\ }\textbf
  {\bibinfo {volume} {123}},\ \bibinfo {pages} {063103} (\bibinfo {year}
  {2018})}\BibitemShut {NoStop}%
\bibitem [{\citenamefont {Cheben}\ \emph {et~al.}(2018)\citenamefont {Cheben},
  \citenamefont {Halir}, \citenamefont {Schmid}, \citenamefont {Atwater},\ and\
  \citenamefont {Smith}}]{cheben2018subwavelength}%
  \BibitemOpen
  \bibfield  {author} {\bibinfo {author} {\bibfnamefont {P.}~\bibnamefont
  {Cheben}}, \bibinfo {author} {\bibfnamefont {R.}~\bibnamefont {Halir}},
  \bibinfo {author} {\bibfnamefont {J.~H.}\ \bibnamefont {Schmid}}, \bibinfo
  {author} {\bibfnamefont {H.~A.}\ \bibnamefont {Atwater}},\ and\ \bibinfo
  {author} {\bibfnamefont {D.~R.}\ \bibnamefont {Smith}},\ }\bibfield  {title}
  {\bibinfo {title} {Subwavelength integrated photonics},\ }\href@noop {}
  {\bibfield  {journal} {\bibinfo  {journal} {Nature}\ }\textbf {\bibinfo
  {volume} {560}},\ \bibinfo {pages} {565} (\bibinfo {year}
  {2018})}\BibitemShut {NoStop}%
\bibitem [{\citenamefont {Joannopoulos}\ \emph {et~al.}(1995)\citenamefont
  {Joannopoulos}, \citenamefont {Meade},\ and\ \citenamefont
  {Winn}}]{bookjoannopoulos}%
  \BibitemOpen
  \bibfield  {author} {\bibinfo {author} {\bibfnamefont {J.~D.}\ \bibnamefont
  {Joannopoulos}}, \bibinfo {author} {\bibfnamefont {R.~D.}\ \bibnamefont
  {Meade}},\ and\ \bibinfo {author} {\bibfnamefont {J.~N.}\ \bibnamefont
  {Winn}},\ }\href@noop {} {\emph {\bibinfo {title} {Photonic Crystals: Molding
  The Flow of Light}}}\ (\bibinfo  {publisher} {Princeton},\ \bibinfo {year}
  {1995})\BibitemShut {NoStop}%
\bibitem [{\citenamefont {Yariv}\ and\ \citenamefont
  {Yeh}(2006)}]{yariv2006photonics}%
  \BibitemOpen
  \bibfield  {author} {\bibinfo {author} {\bibfnamefont {A.}~\bibnamefont
  {Yariv}}\ and\ \bibinfo {author} {\bibfnamefont {P.}~\bibnamefont {Yeh}},\
  }\bibfield  {title} {\bibinfo {title} {Photonics: optical electronics in
  modern communications (the oxford series in electrical and computer
  engineering)},\ }\href@noop {} {\bibfield  {journal} {\bibinfo  {journal}
  {Oxford University Press, Inc}\ }\textbf {\bibinfo {volume} {231}},\ \bibinfo
  {pages} {232} (\bibinfo {year} {2006})}\BibitemShut {NoStop}%
\bibitem [{\citenamefont {Johnson}\ and\ \citenamefont
  {Joannopoulos}(2001)}]{johnson2001block}%
  \BibitemOpen
  \bibfield  {author} {\bibinfo {author} {\bibfnamefont {S.~G.}\ \bibnamefont
  {Johnson}}\ and\ \bibinfo {author} {\bibfnamefont {J.~D.}\ \bibnamefont
  {Joannopoulos}},\ }\bibfield  {title} {\bibinfo {title} {Block-iterative
  frequency-domain methods for {Maxwell’s} equations in a planewave basis},\
  }\href@noop {} {\bibfield  {journal} {\bibinfo  {journal} {Optics Express}\
  }\textbf {\bibinfo {volume} {8}},\ \bibinfo {pages} {173} (\bibinfo {year}
  {2001})}\BibitemShut {NoStop}%
\bibitem [{\citenamefont {Payne}\ and\ \citenamefont
  {Lacey}(1994)}]{payne1994theoretical}%
  \BibitemOpen
  \bibfield  {author} {\bibinfo {author} {\bibfnamefont {F.}~\bibnamefont
  {Payne}}\ and\ \bibinfo {author} {\bibfnamefont {J.}~\bibnamefont {Lacey}},\
  }\bibfield  {title} {\bibinfo {title} {A theoretical analysis of scattering
  loss from planar optical waveguides},\ }\href@noop {} {\bibfield  {journal}
  {\bibinfo  {journal} {Optical and Quantum Electronics}\ }\textbf {\bibinfo
  {volume} {26}},\ \bibinfo {pages} {977} (\bibinfo {year} {1994})}\BibitemShut
  {NoStop}%
\bibitem [{\citenamefont {Grillot}\ \emph {et~al.}(2004)\citenamefont
  {Grillot}, \citenamefont {Vivien}, \citenamefont {Laval}, \citenamefont
  {Pascal},\ and\ \citenamefont {Cassan}}]{grillot2004size}%
  \BibitemOpen
  \bibfield  {author} {\bibinfo {author} {\bibfnamefont {F.}~\bibnamefont
  {Grillot}}, \bibinfo {author} {\bibfnamefont {L.}~\bibnamefont {Vivien}},
  \bibinfo {author} {\bibfnamefont {S.}~\bibnamefont {Laval}}, \bibinfo
  {author} {\bibfnamefont {D.}~\bibnamefont {Pascal}},\ and\ \bibinfo {author}
  {\bibfnamefont {E.}~\bibnamefont {Cassan}},\ }\bibfield  {title} {\bibinfo
  {title} {Size influence on the propagation loss induced by sidewall roughness
  in ultrasmall {SOI} waveguides},\ }\href@noop {} {\bibfield  {journal}
  {\bibinfo  {journal} {IEEE Photonics Technology Letters}\ }\textbf {\bibinfo
  {volume} {16}},\ \bibinfo {pages} {1661} (\bibinfo {year}
  {2004})}\BibitemShut {NoStop}%
\bibitem [{\citenamefont {Ortega-Mo{\~n}ux}\ \emph {et~al.}(2017)\citenamefont
  {Ortega-Mo{\~n}ux}, \citenamefont {{\v{C}}tyrok{\`y}}, \citenamefont
  {Cheben}, \citenamefont {Schmid}, \citenamefont {Wang}, \citenamefont
  {Molina-Fern{\'a}ndez},\ and\ \citenamefont {Halir}}]{ortega2017disorder}%
  \BibitemOpen
  \bibfield  {author} {\bibinfo {author} {\bibfnamefont {A.}~\bibnamefont
  {Ortega-Mo{\~n}ux}}, \bibinfo {author} {\bibfnamefont {J.}~\bibnamefont
  {{\v{C}}tyrok{\`y}}}, \bibinfo {author} {\bibfnamefont {P.}~\bibnamefont
  {Cheben}}, \bibinfo {author} {\bibfnamefont {J.~H.}\ \bibnamefont {Schmid}},
  \bibinfo {author} {\bibfnamefont {S.}~\bibnamefont {Wang}}, \bibinfo {author}
  {\bibfnamefont {{\'I}.}~\bibnamefont {Molina-Fern{\'a}ndez}},\ and\ \bibinfo
  {author} {\bibfnamefont {R.}~\bibnamefont {Halir}},\ }\bibfield  {title}
  {\bibinfo {title} {Disorder effects in subwavelength grating metamaterial
  waveguides},\ }\href@noop {} {\bibfield  {journal} {\bibinfo  {journal}
  {Optics Express}\ }\textbf {\bibinfo {volume} {25}},\ \bibinfo {pages}
  {12222} (\bibinfo {year} {2017})}\BibitemShut {NoStop}%
\bibitem [{\citenamefont {Foster}\ \emph {et~al.}(2004)\citenamefont {Foster},
  \citenamefont {Moll},\ and\ \citenamefont {Gaeta}}]{foster2004optimal}%
  \BibitemOpen
  \bibfield  {author} {\bibinfo {author} {\bibfnamefont {M.~A.}\ \bibnamefont
  {Foster}}, \bibinfo {author} {\bibfnamefont {K.~D.}\ \bibnamefont {Moll}},\
  and\ \bibinfo {author} {\bibfnamefont {A.~L.}\ \bibnamefont {Gaeta}},\
  }\bibfield  {title} {\bibinfo {title} {Optimal waveguide dimensions for
  nonlinear interactions},\ }\href@noop {} {\bibfield  {journal} {\bibinfo
  {journal} {Optics Express}\ }\textbf {\bibinfo {volume} {12}},\ \bibinfo
  {pages} {2880} (\bibinfo {year} {2004})}\BibitemShut {NoStop}%
\bibitem [{\citenamefont {Sato}\ \emph {et~al.}(2015)\citenamefont {Sato},
  \citenamefont {Makino}, \citenamefont {Ishizaka}, \citenamefont {Fujisawa},\
  and\ \citenamefont {Saitoh}}]{sato2015rigorous}%
  \BibitemOpen
  \bibfield  {author} {\bibinfo {author} {\bibfnamefont {T.}~\bibnamefont
  {Sato}}, \bibinfo {author} {\bibfnamefont {S.}~\bibnamefont {Makino}},
  \bibinfo {author} {\bibfnamefont {Y.}~\bibnamefont {Ishizaka}}, \bibinfo
  {author} {\bibfnamefont {T.}~\bibnamefont {Fujisawa}},\ and\ \bibinfo
  {author} {\bibfnamefont {K.}~\bibnamefont {Saitoh}},\ }\bibfield  {title}
  {\bibinfo {title} {A rigorous definition of nonlinear parameter $\gamma$ and
  effective area {$A_{\text{eff}}$} for photonic crystal optical waveguides},\
  }\href@noop {} {\bibfield  {journal} {\bibinfo  {journal} {JOSA B}\ }\textbf
  {\bibinfo {volume} {32}},\ \bibinfo {pages} {1245} (\bibinfo {year}
  {2015})}\BibitemShut {NoStop}%
\bibitem [{\citenamefont {Martini}\ and\ \citenamefont
  {Politi}(2018)}]{martini2018four}%
  \BibitemOpen
  \bibfield  {author} {\bibinfo {author} {\bibfnamefont {F.}~\bibnamefont
  {Martini}}\ and\ \bibinfo {author} {\bibfnamefont {A.}~\bibnamefont
  {Politi}},\ }\bibfield  {title} {\bibinfo {title} {Four wave mixing in {3C}
  {SiC} ring resonators},\ }\href@noop {} {\bibfield  {journal} {\bibinfo
  {journal} {Applied Physics Letters}\ }\textbf {\bibinfo {volume} {112}},\
  \bibinfo {pages} {251110} (\bibinfo {year} {2018})}\BibitemShut {NoStop}%
\bibitem [{\citenamefont {Tan}\ \emph {et~al.}(2010)\citenamefont {Tan},
  \citenamefont {Ikeda}, \citenamefont {Sun},\ and\ \citenamefont
  {Fainman}}]{tan2010group}%
  \BibitemOpen
  \bibfield  {author} {\bibinfo {author} {\bibfnamefont {D.}~\bibnamefont
  {Tan}}, \bibinfo {author} {\bibfnamefont {K.}~\bibnamefont {Ikeda}}, \bibinfo
  {author} {\bibfnamefont {P.}~\bibnamefont {Sun}},\ and\ \bibinfo {author}
  {\bibfnamefont {Y.}~\bibnamefont {Fainman}},\ }\bibfield  {title} {\bibinfo
  {title} {Group velocity dispersion and self phase modulation in silicon
  nitride waveguides},\ }\href@noop {} {\bibfield  {journal} {\bibinfo
  {journal} {Applied Physics Letters}\ }\textbf {\bibinfo {volume} {96}},\
  \bibinfo {pages} {061101} (\bibinfo {year} {2010})}\BibitemShut {NoStop}%
\bibitem [{\citenamefont {Halir}\ \emph {et~al.}(2009)\citenamefont {Halir},
  \citenamefont {Cheben}, \citenamefont {Janz}, \citenamefont {Xu},
  \citenamefont {Molina-Fern{\'a}ndez},\ and\ \citenamefont
  {Wang{\"u}emert-P{\'e}rez}}]{halir2009waveguide}%
  \BibitemOpen
  \bibfield  {author} {\bibinfo {author} {\bibfnamefont {R.}~\bibnamefont
  {Halir}}, \bibinfo {author} {\bibfnamefont {P.}~\bibnamefont {Cheben}},
  \bibinfo {author} {\bibfnamefont {S.}~\bibnamefont {Janz}}, \bibinfo {author}
  {\bibfnamefont {D.-X.}\ \bibnamefont {Xu}}, \bibinfo {author} {\bibfnamefont
  {{\'I}.}~\bibnamefont {Molina-Fern{\'a}ndez}},\ and\ \bibinfo {author}
  {\bibfnamefont {J.~G.}\ \bibnamefont {Wang{\"u}emert-P{\'e}rez}},\ }\bibfield
   {title} {\bibinfo {title} {Waveguide grating coupler with subwavelength
  microstructures},\ }\href@noop {} {\bibfield  {journal} {\bibinfo  {journal}
  {Optics Letters}\ }\textbf {\bibinfo {volume} {34}},\ \bibinfo {pages} {1408}
  (\bibinfo {year} {2009})}\BibitemShut {NoStop}%
\bibitem [{\citenamefont {Cheng}\ \emph {et~al.}(2012)\citenamefont {Cheng},
  \citenamefont {Chen}, \citenamefont {Wong}, \citenamefont {Xu},\ and\
  \citenamefont {Tsang}}]{cheng2012broadband}%
  \BibitemOpen
  \bibfield  {author} {\bibinfo {author} {\bibfnamefont {Z.}~\bibnamefont
  {Cheng}}, \bibinfo {author} {\bibfnamefont {X.}~\bibnamefont {Chen}},
  \bibinfo {author} {\bibfnamefont {C.~Y.}\ \bibnamefont {Wong}}, \bibinfo
  {author} {\bibfnamefont {K.}~\bibnamefont {Xu}},\ and\ \bibinfo {author}
  {\bibfnamefont {H.~K.}\ \bibnamefont {Tsang}},\ }\bibfield  {title} {\bibinfo
  {title} {Broadband focusing grating couplers for suspended-membrane
  waveguides},\ }\href@noop {} {\bibfield  {journal} {\bibinfo  {journal}
  {Optics Letters}\ }\textbf {\bibinfo {volume} {37}},\ \bibinfo {pages} {5181}
  (\bibinfo {year} {2012})}\BibitemShut {NoStop}%
\bibitem [{\citenamefont {Halir}\ \emph {et~al.}(2010)\citenamefont {Halir},
  \citenamefont {Cheben}, \citenamefont {Schmid}, \citenamefont {Ma},
  \citenamefont {Bedard}, \citenamefont {Janz}, \citenamefont {Xu},
  \citenamefont {Densmore}, \citenamefont {Lapointe},\ and\ \citenamefont
  {Molina-Fern{\'a}ndez}}]{halir2010continuously}%
  \BibitemOpen
  \bibfield  {author} {\bibinfo {author} {\bibfnamefont {R.}~\bibnamefont
  {Halir}}, \bibinfo {author} {\bibfnamefont {P.}~\bibnamefont {Cheben}},
  \bibinfo {author} {\bibfnamefont {J.}~\bibnamefont {Schmid}}, \bibinfo
  {author} {\bibfnamefont {R.}~\bibnamefont {Ma}}, \bibinfo {author}
  {\bibfnamefont {D.}~\bibnamefont {Bedard}}, \bibinfo {author} {\bibfnamefont
  {S.}~\bibnamefont {Janz}}, \bibinfo {author} {\bibfnamefont {D.-X.}\
  \bibnamefont {Xu}}, \bibinfo {author} {\bibfnamefont {A.}~\bibnamefont
  {Densmore}}, \bibinfo {author} {\bibfnamefont {J.}~\bibnamefont {Lapointe}},\
  and\ \bibinfo {author} {\bibfnamefont {I.}~\bibnamefont
  {Molina-Fern{\'a}ndez}},\ }\bibfield  {title} {\bibinfo {title} {Continuously
  apodized fiber-to-chip surface grating coupler with refractive index
  engineered subwavelength structure},\ }\href@noop {} {\bibfield  {journal}
  {\bibinfo  {journal} {Optics Letters}\ }\textbf {\bibinfo {volume} {35}},\
  \bibinfo {pages} {3243} (\bibinfo {year} {2010})}\BibitemShut {NoStop}%
\bibitem [{\citenamefont {S{\'a}nchez-Postigo}\ \emph
  {et~al.}(2018)\citenamefont {S{\'a}nchez-Postigo}, \citenamefont
  {Wang{\"u}emert-P{\'e}rez}, \citenamefont {Penad{\'e}s}, \citenamefont
  {Ortega-Mo{\~n}ux}, \citenamefont {Nedeljkovic}, \citenamefont {Halir},
  \citenamefont {Mimun}, \citenamefont {Cheng}, \citenamefont {Qu},
  \citenamefont {Khokhar} \emph {et~al.}}]{sanchez2018mid}%
  \BibitemOpen
  \bibfield  {author} {\bibinfo {author} {\bibfnamefont {A.}~\bibnamefont
  {S{\'a}nchez-Postigo}}, \bibinfo {author} {\bibfnamefont {J.~G.}\
  \bibnamefont {Wang{\"u}emert-P{\'e}rez}}, \bibinfo {author} {\bibfnamefont
  {J.~S.}\ \bibnamefont {Penad{\'e}s}}, \bibinfo {author} {\bibfnamefont
  {A.}~\bibnamefont {Ortega-Mo{\~n}ux}}, \bibinfo {author} {\bibfnamefont
  {M.}~\bibnamefont {Nedeljkovic}}, \bibinfo {author} {\bibfnamefont
  {R.}~\bibnamefont {Halir}}, \bibinfo {author} {\bibfnamefont {F.~E.~M.}\
  \bibnamefont {Mimun}}, \bibinfo {author} {\bibfnamefont {Y.~X.}\ \bibnamefont
  {Cheng}}, \bibinfo {author} {\bibfnamefont {Z.}~\bibnamefont {Qu}}, \bibinfo
  {author} {\bibfnamefont {A.~Z.}\ \bibnamefont {Khokhar}}, \emph {et~al.},\
  }\bibfield  {title} {\bibinfo {title} {Mid-infrared suspended waveguide
  platform and building blocks},\ }\href@noop {} {\bibfield  {journal}
  {\bibinfo  {journal} {IET Optoelectronics}\ }\textbf {\bibinfo {volume}
  {13}},\ \bibinfo {pages} {55} (\bibinfo {year} {2018})}\BibitemShut {NoStop}%
\bibitem [{\citenamefont {S{\'a}nchez-Postigo}\ \emph
  {et~al.}(2019)\citenamefont {S{\'a}nchez-Postigo}, \citenamefont
  {Ortega-Mo{\~n}ux}, \citenamefont {Pereira-Mart{\'\i}n}, \citenamefont
  {Molina-Fern{\'a}ndez}, \citenamefont {Halir}, \citenamefont {Cheben},
  \citenamefont {Penad{\'e}s}, \citenamefont {Nedeljkovic}, \citenamefont
  {Mashanovich},\ and\ \citenamefont
  {Wang{\"u}emert-P{\'e}rez}}]{sanchez2019design}%
  \BibitemOpen
  \bibfield  {author} {\bibinfo {author} {\bibfnamefont {A.}~\bibnamefont
  {S{\'a}nchez-Postigo}}, \bibinfo {author} {\bibfnamefont {A.}~\bibnamefont
  {Ortega-Mo{\~n}ux}}, \bibinfo {author} {\bibfnamefont {D.}~\bibnamefont
  {Pereira-Mart{\'\i}n}}, \bibinfo {author} {\bibfnamefont
  {{\'I}.}~\bibnamefont {Molina-Fern{\'a}ndez}}, \bibinfo {author}
  {\bibfnamefont {R.}~\bibnamefont {Halir}}, \bibinfo {author} {\bibfnamefont
  {P.}~\bibnamefont {Cheben}}, \bibinfo {author} {\bibfnamefont {J.~S.}\
  \bibnamefont {Penad{\'e}s}}, \bibinfo {author} {\bibfnamefont
  {M.}~\bibnamefont {Nedeljkovic}}, \bibinfo {author} {\bibfnamefont
  {G.}~\bibnamefont {Mashanovich}},\ and\ \bibinfo {author} {\bibfnamefont
  {J.}~\bibnamefont {Wang{\"u}emert-P{\'e}rez}},\ }\bibfield  {title} {\bibinfo
  {title} {Design of a suspended germanium micro-antenna for efficient
  fiber-chip coupling in the long-wavelength mid-infrared range},\ }\href@noop
  {} {\bibfield  {journal} {\bibinfo  {journal} {Optics express}\ }\textbf
  {\bibinfo {volume} {27}},\ \bibinfo {pages} {22302} (\bibinfo {year}
  {2019})}\BibitemShut {NoStop}%
\bibitem [{\citenamefont {Krauss}(2007)}]{krauss2007slow}%
  \BibitemOpen
  \bibfield  {author} {\bibinfo {author} {\bibfnamefont {T.~F.}\ \bibnamefont
  {Krauss}},\ }\bibfield  {title} {\bibinfo {title} {Slow light in photonic
  crystal waveguides},\ }\href@noop {} {\bibfield  {journal} {\bibinfo
  {journal} {Journal of Physics D: Applied Physics}\ }\textbf {\bibinfo
  {volume} {40}},\ \bibinfo {pages} {2666} (\bibinfo {year}
  {2007})}\BibitemShut {NoStop}%
\bibitem [{\citenamefont {Rao}\ and\ \citenamefont
  {Hughes}(2007)}]{rao2007single}%
  \BibitemOpen
  \bibfield  {author} {\bibinfo {author} {\bibfnamefont {V.~M.}\ \bibnamefont
  {Rao}}\ and\ \bibinfo {author} {\bibfnamefont {S.}~\bibnamefont {Hughes}},\
  }\bibfield  {title} {\bibinfo {title} {Single quantum-dot {Purcell} factor
  and $\beta$ factor in a photonic crystal waveguide},\ }\href@noop {}
  {\bibfield  {journal} {\bibinfo  {journal} {Physical Review B}\ }\textbf
  {\bibinfo {volume} {75}},\ \bibinfo {pages} {205437} (\bibinfo {year}
  {2007})}\BibitemShut {NoStop}%
\bibitem [{\citenamefont {Arcari}\ \emph {et~al.}(2014)\citenamefont {Arcari},
  \citenamefont {S{\"o}llner}, \citenamefont {Javadi}, \citenamefont {Hansen},
  \citenamefont {Mahmoodian}, \citenamefont {Liu}, \citenamefont {Thyrrestrup},
  \citenamefont {Lee}, \citenamefont {Song}, \citenamefont {Stobbe} \emph
  {et~al.}}]{arcari2014near}%
  \BibitemOpen
  \bibfield  {author} {\bibinfo {author} {\bibfnamefont {M.}~\bibnamefont
  {Arcari}}, \bibinfo {author} {\bibfnamefont {I.}~\bibnamefont {S{\"o}llner}},
  \bibinfo {author} {\bibfnamefont {A.}~\bibnamefont {Javadi}}, \bibinfo
  {author} {\bibfnamefont {S.~L.}\ \bibnamefont {Hansen}}, \bibinfo {author}
  {\bibfnamefont {S.}~\bibnamefont {Mahmoodian}}, \bibinfo {author}
  {\bibfnamefont {J.}~\bibnamefont {Liu}}, \bibinfo {author} {\bibfnamefont
  {H.}~\bibnamefont {Thyrrestrup}}, \bibinfo {author} {\bibfnamefont {E.~H.}\
  \bibnamefont {Lee}}, \bibinfo {author} {\bibfnamefont {J.~D.}\ \bibnamefont
  {Song}}, \bibinfo {author} {\bibfnamefont {S.}~\bibnamefont {Stobbe}}, \emph
  {et~al.},\ }\bibfield  {title} {\bibinfo {title} {Near-unity coupling
  efficiency of a quantum emitter to a photonic crystal waveguide},\
  }\href@noop {} {\bibfield  {journal} {\bibinfo  {journal} {Physical Review
  Letters}\ }\textbf {\bibinfo {volume} {113}},\ \bibinfo {pages} {093603}
  (\bibinfo {year} {2014})}\BibitemShut {NoStop}%
\bibitem [{\citenamefont {Nagy}\ \emph {et~al.}(2019)\citenamefont {Nagy},
  \citenamefont {Niethammer}, \citenamefont {Widmann}, \citenamefont {Chen},
  \citenamefont {Udvarhelyi}, \citenamefont {Bonato}, \citenamefont {Hassan},
  \citenamefont {Karhu}, \citenamefont {Ivanov}, \citenamefont {Son} \emph
  {et~al.}}]{nagy2019high}%
  \BibitemOpen
  \bibfield  {author} {\bibinfo {author} {\bibfnamefont {R.}~\bibnamefont
  {Nagy}}, \bibinfo {author} {\bibfnamefont {M.}~\bibnamefont {Niethammer}},
  \bibinfo {author} {\bibfnamefont {M.}~\bibnamefont {Widmann}}, \bibinfo
  {author} {\bibfnamefont {Y.-C.}\ \bibnamefont {Chen}}, \bibinfo {author}
  {\bibfnamefont {P.}~\bibnamefont {Udvarhelyi}}, \bibinfo {author}
  {\bibfnamefont {C.}~\bibnamefont {Bonato}}, \bibinfo {author} {\bibfnamefont
  {J.~U.}\ \bibnamefont {Hassan}}, \bibinfo {author} {\bibfnamefont
  {R.}~\bibnamefont {Karhu}}, \bibinfo {author} {\bibfnamefont {I.~G.}\
  \bibnamefont {Ivanov}}, \bibinfo {author} {\bibfnamefont {N.~T.}\
  \bibnamefont {Son}}, \emph {et~al.},\ }\bibfield  {title} {\bibinfo {title}
  {High-fidelity spin and optical control of single silicon-vacancy centres in
  silicon carbide},\ }\href@noop {} {\bibfield  {journal} {\bibinfo  {journal}
  {Nature communications}\ }\textbf {\bibinfo {volume} {10}} (\bibinfo {year}
  {2019})}\BibitemShut {NoStop}%
\bibitem [{\citenamefont {Alferness}(1982)}]{alferness1982waveguide}%
  \BibitemOpen
  \bibfield  {author} {\bibinfo {author} {\bibfnamefont {R.~C.}\ \bibnamefont
  {Alferness}},\ }\bibfield  {title} {\bibinfo {title} {Waveguide electrooptic
  modulators},\ }\href@noop {} {\bibfield  {journal} {\bibinfo  {journal} {IEEE
  Transactions on Microwave Theory Techniques}\ }\textbf {\bibinfo {volume}
  {30}},\ \bibinfo {pages} {1121} (\bibinfo {year} {1982})}\BibitemShut
  {NoStop}%
\bibitem [{\citenamefont {Wang}\ \emph {et~al.}(2018)\citenamefont {Wang},
  \citenamefont {Zhang}, \citenamefont {Stern}, \citenamefont {Lipson},\ and\
  \citenamefont {Lon{\v{c}}ar}}]{wang2018nanophotonic}%
  \BibitemOpen
  \bibfield  {author} {\bibinfo {author} {\bibfnamefont {C.}~\bibnamefont
  {Wang}}, \bibinfo {author} {\bibfnamefont {M.}~\bibnamefont {Zhang}},
  \bibinfo {author} {\bibfnamefont {B.}~\bibnamefont {Stern}}, \bibinfo
  {author} {\bibfnamefont {M.}~\bibnamefont {Lipson}},\ and\ \bibinfo {author}
  {\bibfnamefont {M.}~\bibnamefont {Lon{\v{c}}ar}},\ }\bibfield  {title}
  {\bibinfo {title} {Nanophotonic lithium niobate electro-optic modulators},\
  }\href@noop {} {\bibfield  {journal} {\bibinfo  {journal} {Optics express}\
  }\textbf {\bibinfo {volume} {26}},\ \bibinfo {pages} {1547} (\bibinfo {year}
  {2018})}\BibitemShut {NoStop}%
\bibitem [{\citenamefont {Janner}\ \emph {et~al.}(2009)\citenamefont {Janner},
  \citenamefont {Tulli}, \citenamefont {Garc{\'\i}a-Granda}, \citenamefont
  {Belmonte},\ and\ \citenamefont {Pruneri}}]{janner2009micro}%
  \BibitemOpen
  \bibfield  {author} {\bibinfo {author} {\bibfnamefont {D.}~\bibnamefont
  {Janner}}, \bibinfo {author} {\bibfnamefont {D.}~\bibnamefont {Tulli}},
  \bibinfo {author} {\bibfnamefont {M.}~\bibnamefont {Garc{\'\i}a-Granda}},
  \bibinfo {author} {\bibfnamefont {M.}~\bibnamefont {Belmonte}},\ and\
  \bibinfo {author} {\bibfnamefont {V.}~\bibnamefont {Pruneri}},\ }\bibfield
  {title} {\bibinfo {title} {Micro-structured integrated electro-optic
  {LiNbO$_3$} modulators},\ }\href@noop {} {\bibfield  {journal} {\bibinfo
  {journal} {Laser \& Photonics Reviews}\ }\textbf {\bibinfo {volume} {3}},\
  \bibinfo {pages} {301} (\bibinfo {year} {2009})}\BibitemShut {NoStop}%
\bibitem [{\citenamefont {Xu}\ \emph {et~al.}(2005)\citenamefont {Xu},
  \citenamefont {Schmidt}, \citenamefont {Pradhan},\ and\ \citenamefont
  {Lipson}}]{xu2005micrometre}%
  \BibitemOpen
  \bibfield  {author} {\bibinfo {author} {\bibfnamefont {Q.}~\bibnamefont
  {Xu}}, \bibinfo {author} {\bibfnamefont {B.}~\bibnamefont {Schmidt}},
  \bibinfo {author} {\bibfnamefont {S.}~\bibnamefont {Pradhan}},\ and\ \bibinfo
  {author} {\bibfnamefont {M.}~\bibnamefont {Lipson}},\ }\bibfield  {title}
  {\bibinfo {title} {Micrometre-scale silicon electro-optic modulator},\
  }\href@noop {} {\bibfield  {journal} {\bibinfo  {journal} {Nature}\ }\textbf
  {\bibinfo {volume} {435}},\ \bibinfo {pages} {325} (\bibinfo {year}
  {2005})}\BibitemShut {NoStop}%
\end{thebibliography}%

\end{document}